\documentclass[graybox]{SNmult}

\usepackage{type1cm}        
\usepackage{makeidx}         
\usepackage{graphicx}        
\usepackage{multicol}        
\usepackage[bottom]{footmisc}
\usepackage{amsmath, amssymb}
\usepackage{xcolor}
\usepackage{subcaption}

\usepackage{booktabs}
\usepackage{tabularx}
\usepackage{array}

\usepackage{newtxtext}       %
\usepackage[varvw]{newtxmath}       

\usepackage{amsthm}
\theoremstyle{definition} 

\theoremstyle{remark} 

\makeindex             

\newcommand{\Na}{{N_{\rm a}}}
\newcommand{\No}{{N_{\rm o}}}

\newcommand{\Aa}{A_{\rm a}}

\begin{document}
\title*{Tunable Spatial Task Allocation with Application to Maritime Swarms}
\author{Vishnudatta Thota\orcidID{0009-0003-1297-7978}, \\
Tyler Paine\orcidID{0000-0001-6071-621X} and, \\ 
Anastasia Bizyaeva\orcidID{0000-0003-0728-9655}}
\institute{Vishnudatta Thota \at Sibley School of Mechanical and Aerospace Engineering, Cornell University, Ithaca, NY, USA, 14853, \email{vt279@cornell.edu}
\and Tyler Paine \at Department of Mechanical Engineering, Massachusetts Institute of Technology, Cambridge, MA, USA, \email{tpaine@mit.edu}
\and Anastasia Bizyaeva \at Sibley School of Mechanical and Aerospace Engineering, Cornell University, Ithaca, NY, USA, 14853, \email{anastasiab@cornell.edu}}
%
%
\maketitle
\abstract*{ 
Each chapter should be preceded by an abstract (no more than 200 words) that summarizes the content. The abstract will appear \textit{online} 
at \verb|www.SpringerLink.com| and be available with unrestricted access. This allows unregistered users to read the abstract as a teaser for the complete chapter.
Please use the \verb|abstract*| command for typesetting the text of the online abstracts (cf. source file of this chapter template abstract) and include them with the source files of your manuscript. Use the plain \verb|abstract| command if the abstract is also to appear in the printed version of the book.
}

\abstract{In this chapter, we explore how a team of autonomous agents can flexibly reconfigure the distribution of its members across many spatially embedded tasks using a task allocation strategy grounded in a generalization of the recently proposed Nonlinear Opinion Dynamics model. Motivated by an adaptive maritime survey application, we couple opinion dynamics directly to spatial movement and examine how the emergent macroscopic allocation is shaped by the interplay between environmental geometry, nature of social interactions, and task demands. We illustrate in simulation how social coupling in the model tunes group cohesion, producing macroscopic allocation patterns ranging from maximally dispersed states, to intermediate semi-dispersed states with a smaller number of distinct subgroups, to maximally cohesive consensus states, in distinct and interpretable parameter regimes. Furthermore, we demonstrate with illustrative examples how a team can reconfigure itself in real time under changing task demands. For allocation scenarios with distance-based task preference and matched number of agents and tasks, we relate numerically the final allocation in the group to the Hungarian optimum, illustrating close-to-optimal performance. We illustrate performance of this method under realistic communication constraints in the MOOS-IvP-Multi simulation environment with simulated autonomous underwater vehicles (AUVs). Together, these examples show how a common set of local interaction rules can support a large range of interpretable macroscopic allocation states and flexible transitions between them. 
}

\section{Introduction}
\label{sec:Introduction}

Allocating members of an autonomous team across tasks flexibly without centralized control is a key challenge in multi-agent autonomy. For example, a team of robots surveying an environment may need to distribute its members across candidate sites, then form subgroups to investigate selected sites where mission-relevant information was located, and finally reallocate robots to other sites as tasks within the initially selected sites are completed. Each of the three mission stages requires individual robots to select specific sites to commit to based on available information, but they each require different macroscopic patterns of collective participation within the group. Furthermore, the evolving mission scenario requires the group to transition between different coalitions as task-relevant information evolves. When tasks are embedded in physical space, as they frequently are in robotic applications, the interplay between the geometry of the task distribution and the agents' physical motion also plays a key role in the emergent decision-making in the group. These considerations motivate the need for decentralized task allocation frameworks that account for the coupling between decision-making and physical motion, and also provide interpretable mechanisms to shape collective patterns of participation through tunable local interactions. In this chapter, we illustrate how nonlinear opinion dynamics \cite{bizyaeva2023tac} can support such flexible allocation across many spatial tasks, with a social coupling parameter providing an interpretable design ``knob'' that tunes the degree of cohesion within a group.

The multi-robot task allocation (MRTA) problem is commonly modeled as a combinatorial optimization problem that accounts for task utilities and constraints \cite{gerkey2004formal}. Decentralized MRTA approaches prominently include bio-inspired threshold mechanisms based on local responses to sensory stimulus \cite{agassounon2002efficiency,kalra2006comparative,castello2016adaptive} and market-based approaches in which agents bid on tasks through repeated rounds of auctions \cite{choi2009consensus,quinton2023market}. Another class of approaches associates task preferences of each agent with dynamic variables such as \textit{opinions} or \textit{motivations} that are exchanged over a communication network. This perspective transforms the task allocation problem into a continuous-time dynamical system convergence problem. There is a long precedent for using opinion dynamics models as coordination mechanisms for collective choice; for example, in \cite{montes2011majority} a majority-rule opinion dynamics model is used for collective decision-making between two alternative travel paths. In more recent works, vector-valued opinion and motivation dynamics are used to coordinate assignments among multiple tasks \cite{reverdy2021motivation,thompson2021multi,baxevani2023multi,zhang2024opinion,zhang2025opinion}. The dynamical approach provides a natural framework for studying how decision rules, interactions among agents, and local information shape collective allocation patterns. 

Recently proposed \textit{Nonlinear Opinion Dynamics (NOD)} provide a tractable mathematical framework for relating patterns of collective decisions in a group to values of social interaction parameters, network structure, and influence of external information \cite{bizyaeva2023tac, bizyaeva2025tac,leonard2024fast}. In this framework, each agent's opinions evolve through a bounded nonlinear response to a combination of self-reinforcement, social influence, and external inputs. The model form takes direct inspiration from dynamics of influence and decision-making over social networks. The nonlinearity of the social interactions enables multistable opinion patterns and gives rise to bifurcations through which the number or stability of these macroscopic patterns changes. Interpretable parameters in this framework, such as the signs of pairwise social couplings and local biases, can then be used to shape the emergent allocation as has been illustrated for two-task decentralized allocation scenarios \cite{bizyaeva2022switching,amorim2024threshold}. Recently, NOD has been applied to the problem of autonomous teaming within the context of a hierarchical coordination framework \cite{paine2024model,paine2025census}. However, most practical scenarios considered in the listed works are still limited to allocation of agents to two or three subgroups. Our goal in this chapter is to illustrate how agents can flexibly allocate themselves across a large number of spatially embedded tasks using the NOD framework as an organizing principle.

When deciding between multiple tasks, it is helpful to distinguish between two desirable aspects of the decision process. Within each agent, dynamics of opinions must resolve competition among feasible options, i.e. one opinion must dominate each agent's preferences to avoid internal deadlocks. Across agents, the interactions must organize resulting preferences into a macroscopic pattern of commitment or allocation, shaped by the nature of social interactions between agents and coupled to the geometry of tasks and local information about them. In the original NOD formulation \cite{bizyaeva2023tac} and its subsequent extensions and applications, nonlinear interactions between agents are modeled as componentwise saturations, with a typical choice of nonlinearity being the hyperbolic tangent function. This is a natural nonlinearity for the description of binary choice between mutually exclusive tasks; however, in multi-task settings, saturation alone does not ensure that one task preference strongly dominates over others. In this chapter, we utilize an NOD formulation with a shifted softmax saturation of social information that more naturally represents choice between multiple mutually exclusive options, and explicitly connects the mathematical structure of the NOD framework with established game-theoretic and neural computation frameworks for decision-making and task allocation, including logit population game dynamics \cite{sandholm2010population,barreiro2016distributed,cianfanelli2025stability} and Potts neural network models \cite{hakkinen2000local,david2022deterministic}.

Several previous works have established connections between NOD and game theoretic frameworks. For example, \cite{park2021tuning} uses NOD to tune cooperative behavior in social dilemma games through a competition between a payoff function and social influence; \cite{paine2024adaptive} uses a game-theoretic formulation to tune a bias in NOD to reallocate a team between two tasks. More recently \cite{sinhmar2026neuromorphic} shows how ring attractor-like internal coupling among an agent's preferences towards actions in an NOD framework shapes a neuromorphic best-response mechanism, with an application to a multi-agent coverage game with no explicit communication. 
Building on these connections, we study how social interactions, spatial movement, and task capacity demands jointly shape allocation across many spatially embedded tasks in a softmax NOD formulation that is coupled to physical motion of agents in the plane.

To illustrate how softmax NOD can be leveraged in flexible spatial task allocation scenarios, we study a coupled model of opinion dynamics and physical movement, with an adaptive maritime survey serving as a motivating application throughout the chapter. The chapter is structured the following way. In Section~\ref{sec:Motivating Example} we introduce the motivating scenario. In Section~\ref{sec:Problem Setup} we present the softmax NOD task allocation model, introducing agent-specific task capabilities as a mask on the nonlinearity, and discuss the connection between the resulting model and logit dynamics. We illustrate how decisiveness emerges in a bifurcation of a neutral opinion state. In Section~\ref{sec:Behavioral regimes of the coupled dynamics} we present numerical studies that examine how social coupling shapes the collective allocation, illustrating that a single parameter that tunes the level of cooperation or antagonism in the group triggers a sequence of transitions in the macroscopic allocation state of the group, from maximally dispersed in the limit of strong antagonism, to a consensus on a common aggregation site in the limit of strong cooperativity. In scenarios with equal numbers of agents and tasks, we compare the maximally dispersed regime to the Hungarian optimum, investigate incomplete communication graphs, and illustrate in simulation how a group can redistribute itself when task capacities change. In Section~\ref{sec:Simulations} we translate this framework to a simulation environment for autonomous underwater vehicles in MOOS-IvP-Multi \cite{benjamin2010MOOSIvP} with limited communication range and discretized opinion exchanges, illustrating performance of this task allocation strategy under more realistic operating conditions. We conclude in Section~\ref{sec:Concluding Remarks}.

\section{Motivating Example: Adaptive Survey of Benthic Fauna}
\label{sec:Motivating Example}
We consider the class of problems that require dynamic task allocation under significant communication constraints.   The following oceanographic survey scenario grounds the theoretical development in a practical application. 

The challenge we address is locating and quantifying adult benthic fauna that are found on the ocean floor at depths of approximately 100-300 m or in the deep water column near the ocean floor.  The specific organism of interest is \textit{Pleuroncodes planipes}, or the tuna crab, which is commonly found along the Pacific coast of North and South America \cite{boyd1967benthic}.  Recently, large swarms of \textit{P. planipes} were discovered on the ocean floor \cite{pineda2016crab}, motivating further research into the behavior of these organisms.

\textit{P. planipes} forms a key link in the food chain by connecting smaller phytoplankton to larger predators such as tuna in upwelling-driven ecosystems \cite{alverson1963food}.  These upwelling systems are highly productive fisheries along the coasts of California, Mexico, and Peru. Since they play a central role in the ecosystem, it is important to understand their spatial distribution, particularly that of the adult benthic \textit{P. planipes}, which are believed to be responsible for the majority of reproduction \cite{longhurst1971breeding}.  

Despite their ecological importance, these organisms are difficult to survey. \textit{P. planipes} migrate as larvae or juveniles suspended in pelagic water column. The migration patterns are driven by ocean currents, and blooms are correlated with upwelling \cite{Robinson2004Upwelling}. Both of life phases are affected by processes that are altered by ocean warming.  Adults are no easier to observe. Mature adults are mostly found at depths of 100-300 m \cite{boyd1967benthic}, preventing remote sensing via satellites, while costly trawling surveys and hydroacoustic mapping provide only an estimate of biomass in the water column \cite{Robinson2004Upwelling}. Moreover, adults can swarm together to form tight clusters on the ocean floor, or otherwise exhibit high spatial variability \cite{pineda2016crab}. 

One viable approach is to use autonomous vehicles that descend into the water column and illuminate an area of ocean floor to capture images and samples \cite{pineda2016crab}.  There are several autonomous underwater vehicles (AUVs) in service that already perform missions of this type \cite{singh2002seabed, Kaiser2016Sentry, Jones2014Slocum}, but they are often expensive and require dedicated crewed ships to support operations, limiting their ability to perform large-scale, persistent searches. Recently, lower-cost vehicles have become more capable and offer an affordable alternative platform for surveying at 300 m \cite{Phillips2017EcoSub,Turrisi2024Spurdog,Schneider2024Jaia}.  Although these low-cost vehicles are less ruggedized, they rely on more sophisticated autonomy and control to operate at the edge of their relatively limited operating envelope.  Their low cost also allows them to be deployed in groups, further improving robustness and efficiency. 

The search area is simply too large for an exhaustive search.   Even with a good prior estimate of the location of the benthic fauna, the search area is on the order of 100 $km^2$\cite{boyd1967benthic}. Low-cost underwater vehicles have a range of 4-5 km and carry cameras that image areas on the order of 10$m^2$ per frame. In addition, vehicles do not have the communication capacity or computing resources to solve assignment problems using brute-force approaches, especially as team size grows. 

\begin{figure}[ht]
  \centering
  \begin{subfigure}[t]{0.48\textwidth}
    \includegraphics[width=\linewidth]{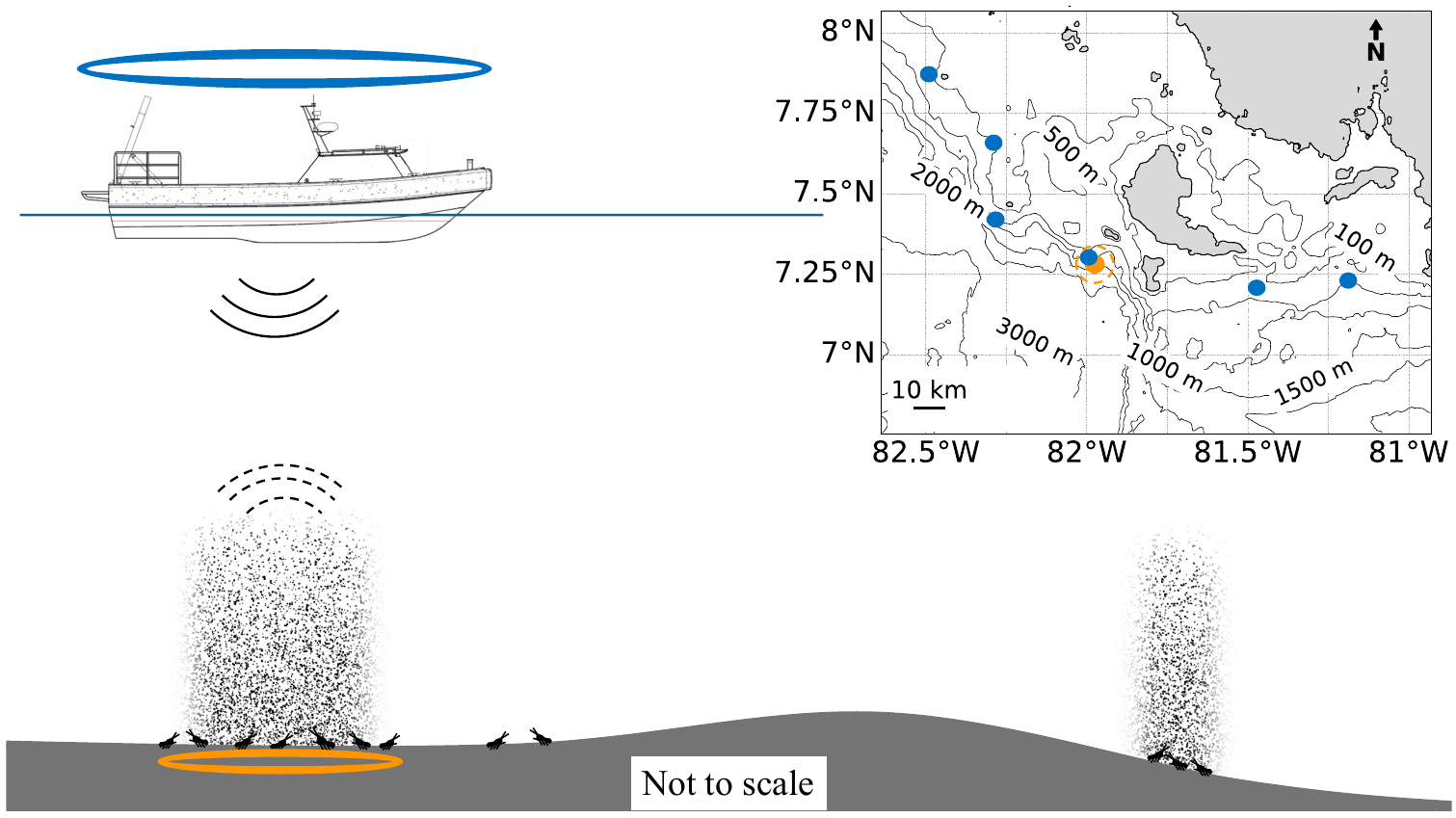}
    \caption{}
    \label{fig:ex_step_1}
  \end{subfigure}\hfill
  \begin{subfigure}[t]{0.48\textwidth}
    \includegraphics[width=\linewidth]{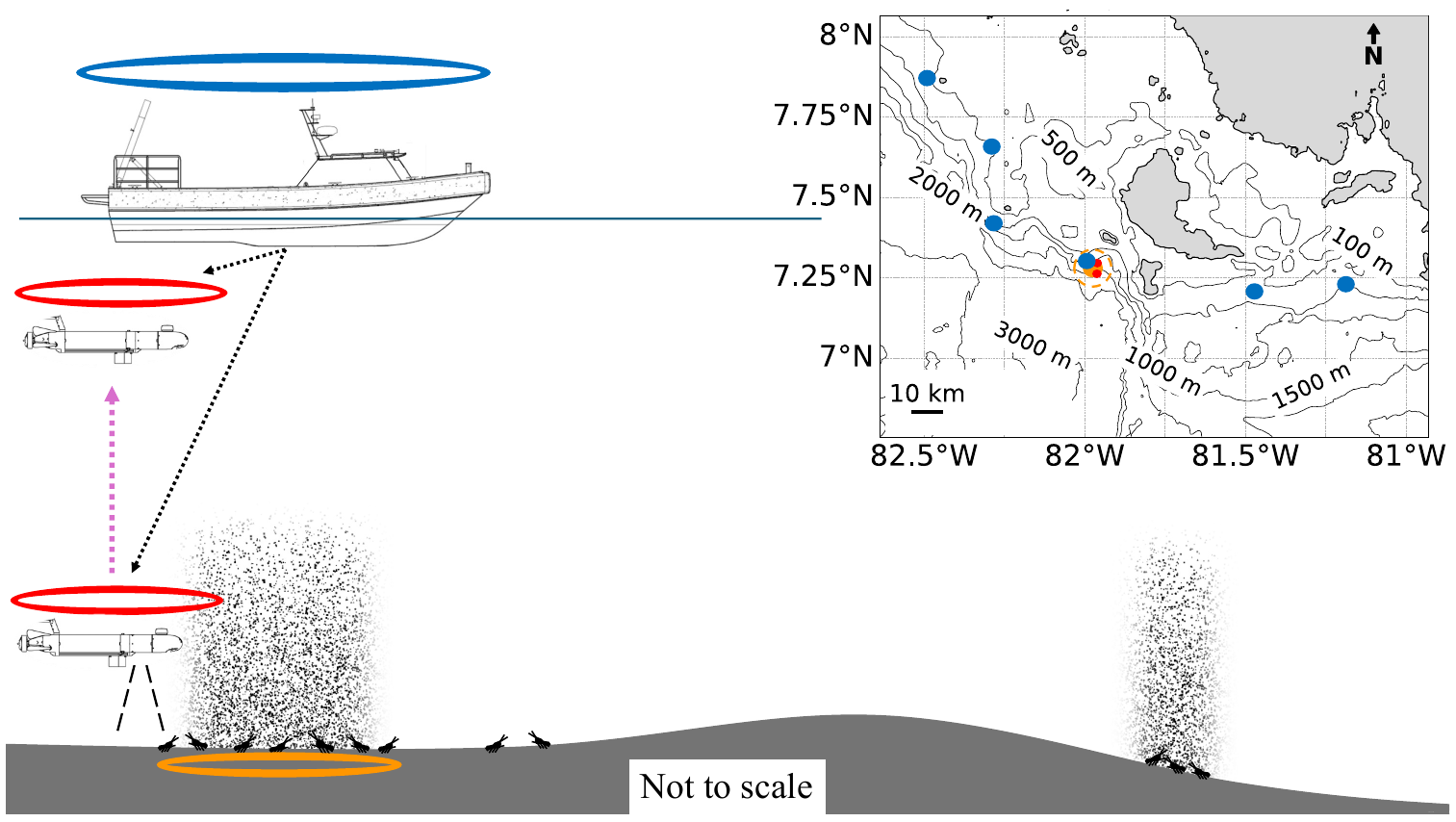}
    \caption{}
    \label{fig:ex_step_2}
  \end{subfigure}

  \vspace{6pt}

  \begin{subfigure}[t]{0.48\textwidth}
    \includegraphics[width=\linewidth]{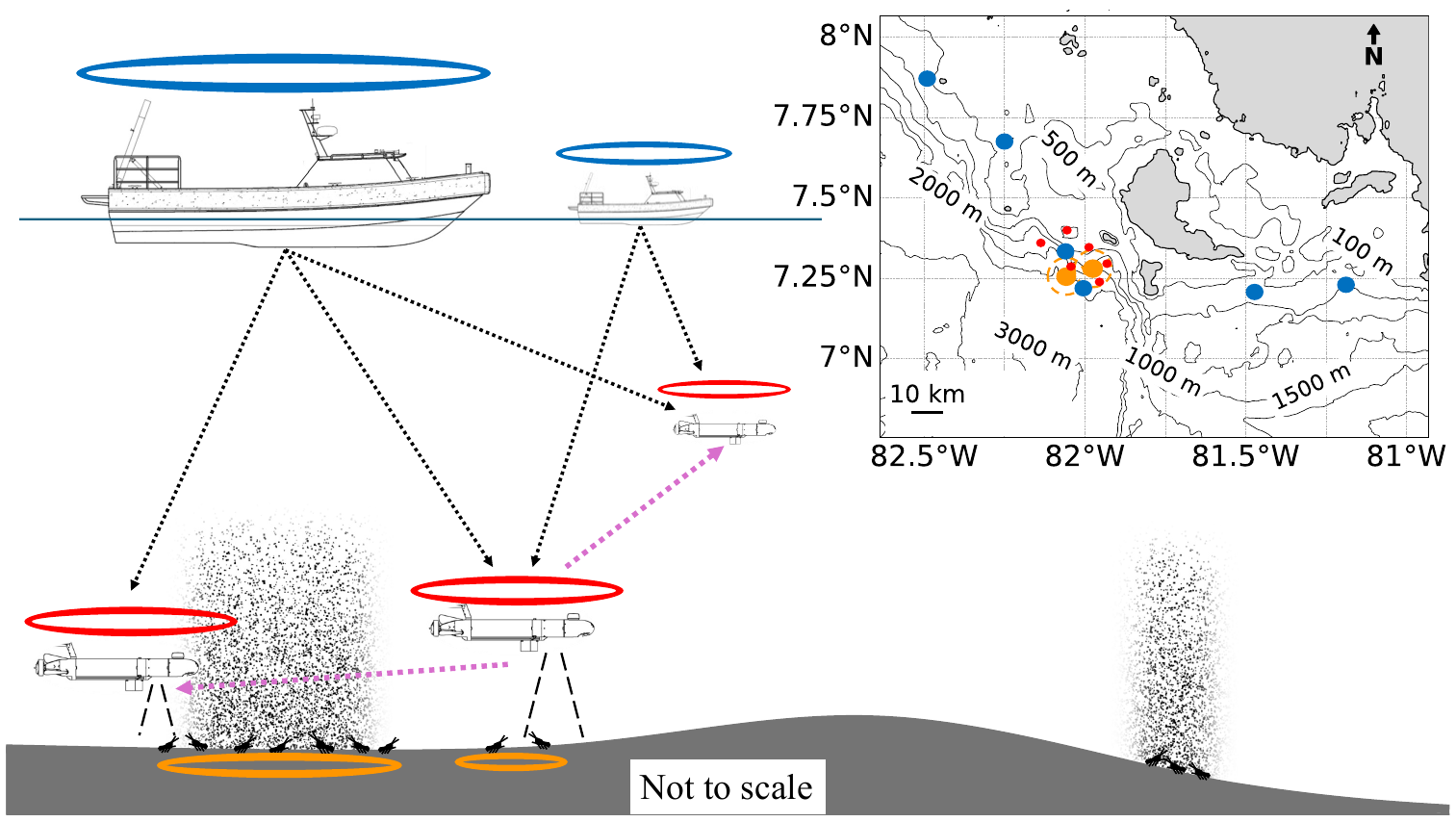}
    \caption{}
    \label{fig:ex_step_3}
  \end{subfigure}\hfill
  \begin{subfigure}[t]{0.48\textwidth}
    \includegraphics[width=\linewidth]{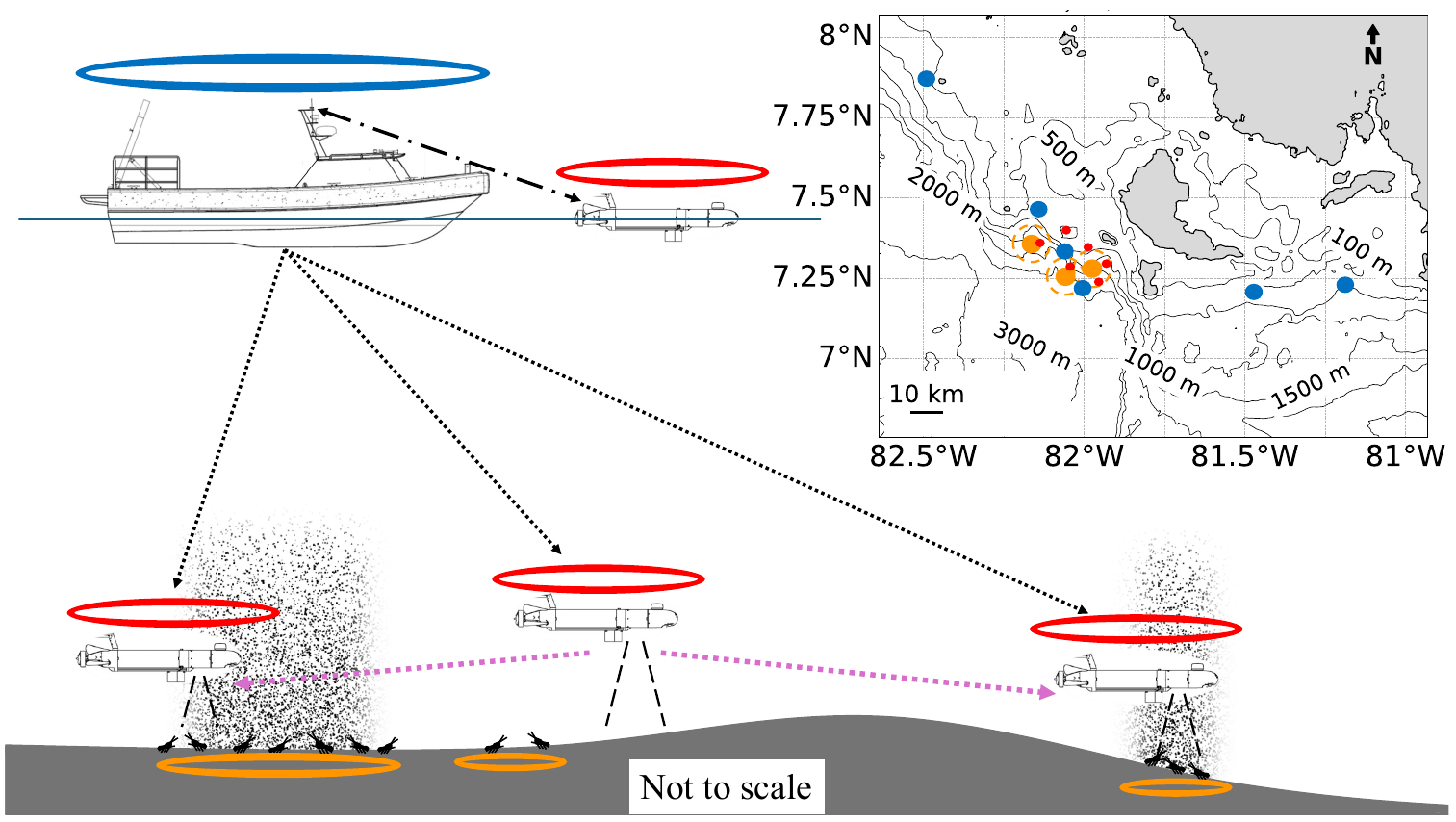}
    \caption{}
    \label{fig:ex_step_4}
  \end{subfigure}

  \Description{Four-panel concept of operations diagram. (a) USVs survey an area
  using hydroacoustic backscatter to detect adult tuna crabs in the water column.
  (b) A USV detects mass in the water column and deploys AUVs to investigate,
  providing one-way travel time messages with GPS information; the AUVs broadcast
  messages between each other without confirmation. (c) Other USVs join the search
  and deploy their own AUVs, which work together to sample existing crab clusters
  or search for nearby ones. (d) A low-battery AUV surfaces to transmit data to the
  USVs, and is recovered and recharged for another deployment.}

  \caption{Concept of operations for hypothetical cooperative USV--AUV surveying of tuna crab
  aggregations. \textbf{(a)} USVs (blue) survey the area using hydroacoustic
  backscatter surveys to detect adult tuna crabs in the water column.
  \textbf{(b)} A USV detects mass in the water column and deploys AUVs (red) to
  investigate, providing one-way travel time messages with GPS information; the
  AUVs communicate between each other with a broadcast message (orange) without
  confirmation. \textbf{(c)} Other USVs decide to help search the area and deploy
  their AUVs, which work together to either sample an existing cluster of crabs or
  search for nearby clusters. \textbf{(d)} When an AUV is almost out of battery
  power it surfaces to transmit data to the USV(s), which recovers the AUV if
  possible and recharges it for another deployment. 
  }
  \label{fig:ex_steps}
\end{figure}

One way to address this problem is to use uncrewed surface vehicles (USVs) to perform a preliminary search using acoustic backscatter methods.  Each USV also carries a complement of AUVs that it can deploy. Both advantages of using USVs stem from energy savings: the USVs can efficiently refine the search area by transiting on the surface, and then they transport the energy-constrained AUVs as close to the estimated sample sites as possible.
Although this multi-robot system design relies on largely commercially available hardware, new distributed autonomy techniques are required to coordinate these teams at scale.  Fig.~\ref{fig:ex_steps} illustrates the proposed survey scenario and the associated stages of cooperative task allocation.

\section{Problem Setup}
\label{sec:Problem Setup}
In this section, we introduce notation from graph theory and present the coupled agent opinion, motion, and task price dynamics that comprise the task allocation framework studied in the rest of the chapter.

\subsection{Preliminaries}
An unweighted, unsigned graph $\mathcal{G} = (\mathcal{V},\mathcal{E},A)$ consists of a node set $\mathcal{V} = \{1, \dots, n\}$, an edge set $\mathcal{E}\subseteq \mathcal{V} \times \mathcal{V}$, and adjacency matrix $A \in \mathbb{R}^{n \times n}$ with entries $a_{ij} =1$ if $(i,j) \in \mathcal{E}$ and $a_{ij} = 0$ otherwise. We consider simple graphs, i.e., ones with no self-loops, $a_{ii} = 0$ for all $i \in \mathcal{V}$. 

A graph is undirected if $(i,j) \in \mathcal{E}$ if and only if $(j,i) \in \mathcal{E}$, which is equivalent to $A = A^T$. A graph is connected if there exists a path between every pair of distinct nodes. 
A graph is complete if there is an edge between every pair of distinct nodes. 

We also use the masked softmax function in this chapter, which assigns nonzero weights only to the indices in a specified set $\mathcal{A}$.
It is defined as follows:
\begin{equation}
\label{eq:masked_softmax_definition}
\big[\sigma_{\mathcal{A}}(\mathbf{w})\big]_j =
\begin{cases}
    \dfrac{e^{w_j}}{\sum_{l \in \mathcal{A}} e^{w_l}}, & j \in \mathcal{A},\\[2mm]
    0, & j \notin \mathcal{A},
\end{cases},
\end{equation}

We summarize the notations used in this chapter in Table~\ref{tab:notation_table}. Vectors are denoted by bold letters.

\begin{table*}[t]
    \centering
    \caption{Summary of notations used in this chapter.}
    \label{tab:notation_table}
    \small
    \renewcommand{\arraystretch}{1.15}
    \begin{tabularx}{\textwidth}{
        >{\centering\arraybackslash}p{0.25\textwidth}
        >{\centering\arraybackslash}p{0.20\textwidth}
        X
    }
        \toprule
        \textbf{Notation} & \textbf{Space} & \textbf{Description} \\
        \midrule

        $\Na$ & $\mathbb{N}$ & Number of agents. \\

        $\No$ & $\mathbb{N}$ & Number of sampling sites (options). \\

        $\mathcal{G}$ & -- & Communication graph of agents. \\

        $\mathcal{V}$ & -- & Node set of agents. \\

        $\mathcal{E}$ & -- & Edge set of agents. \\

        $\Aa$ & $\{0,1\}^{\Na \times \Na}$ & Adjacency matrix of the communication graph. \\

        $\mathbf{X}_i$ & $\mathbb{R}^{2}$ & Physical position of agent $i$. \\

        $\mathbf{X}_{Gj}$ & $\mathbb{R}^{2}$ & Physical position of sampling site $j$. \\

        $k$ & $\mathbb{R}^{\geq 0}$ & Spatial control gain. \\

        $\beta_{x}$ & $\mathbb{R}^{\geq 0}$ & Position dynamics sensitivity. \\

        $\mathbf{Z}_i = (z_{i1},\ldots,z_{i\No})$ & $\mathbb{R}^{\No}$ & Opinion vector of agent $i$. \\

        $z_{ij}$ & $\mathbb{R}$ & Opinion of agent $i$ regarding sampling site $j$. \\

        $\mathcal{A}_{i}$ & -- & Capability set of agent $i$. \\

        $\sigma_{\mathcal{A}_i}$ & -- & Masked softmax function over capability set $\mathcal{A}_{i}$. \\

        $\kappa$ & $\mathbb{R}^{\geq 0}$ & Bias adaptation rate. \\

        $v_{\max}$ & $\mathbb{R}^{\geq 0}$ & Agent saturation speed. \\

        $\beta_{s}$ & $\mathbb{R}^{\geq 0}$ & Opinion saturation sensitivity. \\

        $|\mathcal{A}_i|$ & $\mathbb{N}$ & Cardinality of capability set $\mathcal{A}_i$. \\

        $\mathbf{1}_{\mathcal{A}_i}$ & $\{0,1\}^{\No}$ & Indicator vector of capability set $\mathcal{A}_i$. \\

        $\mathbf{b}_i = (b_{i1},\ldots,b_{i\No})$ & $\mathbb{R}^{\No}$ & Bias vector of agent $i$. \\

        $b_{ij}$ & $\mathbb{R}$ & Bias of agent $i$ regarding sampling site $j$. \\

        $u$ & $\mathbb{R}^{\geq 0}$ & Attention parameter. \\

        $d$ & $\mathbb{R}^{\geq 0}$ & Damping coefficient. \\

        $\gamma$ & $\mathbb{R}$ & Social influence weight. \\

        $\alpha$ & $\mathbb{R}^{\geq 0}$ & Opinion self-reinforcement strength. \\

        $\tau_{z}$ & $(0,1)$ & Opinion dynamics timescale constant. \\

        $c_{j}$ & $\mathbb{W}$ & Capacity of sampling site $j$. \\

        $\lambda_{j}$ & $\mathbb{R}_{\geq 0}$ & Dynamic penalty of over-subscribed site $j$. \\

        $\beta_{l}$ & $\mathbb{R}^{\geq 0}$ & Sharpness for the price-mediated capacity feedback. \\

        $\epsilon$ & $(0,1)$ & Price adaptation rate. \\

        $\mathbf{0}_{\No}$ & $\mathbb{R}^{\No}$ & All zeros vector. \\

        $\mathbf{1}_{\No}$ & $\mathbb{R}^{\No}$ & All ones vector. \\

        $I$ & -- & Identity matrix of suitable dimension. \\

        \bottomrule
    \end{tabularx}
\end{table*}

\subsection{Agent dynamics}
\label{subsec:Agent dynamics}
We consider a group of $\Na$ agents that must collectively select among $\No$ spatially distributed options, or sampling sites. Each agent's opinion vector represents its preference for the corresponding option. The agents exchange their opinions over an undirected, connected, unweighted, unsigned communication graph while their physical motion is driven by their current opinions of the sampling sites. The agent opinions are biased by their distance from the sampling sites. This formulation therefore couples the opinion dynamics and the physical movement of the agents. 

Let $\mathbf{X}_i \in \mathbb{R}^2$ denote the position of agent $i \in \{1,\cdots,\Na\}$, and let $\mathbf{X}_{Gj} \in \mathbb{R}^2$, denote the fixed locations of the $\No$ spatially distributed options. Each agent maintains an opinion vector
\begin{equation}
\mathbf{Z}_{i} = (z_{i1},\dots,z_{i\No})^{\top} \in \mathbb{R}^{\No},
\end{equation}
where $z_{ij}$ represents the agent's preference for option $j \in \{1,\cdots,\No\}$. 
The interaction among the agents is represented by the graph $\mathcal{G} = (\mathcal{V},\mathcal{E},\Aa)$, where node set $\mathcal{V} = \{1, \dots, \Na\}$, an edge set $\mathcal{E}\subseteq \mathcal{V} \times \mathcal{V}$, and adjacency matrix $\Aa \in \mathbb{R}^{\Na \times \Na}$ with entries $(\Aa)_{ij} =1$ if $(i,j) \in \mathcal{E}$ and $(\Aa)_{ij} = 0$ otherwise. 

Depending on the initial positions of the agents or their physical limitations, not all the agents can go to all the sample sites. Let capability set 
\begin{equation}
\mathcal{A}_i \subseteq \{1,\dots,\No\}
\end{equation}
denote the set of options the agent $i$ is capable of performing. 

Spatial information enters the opinion dynamics through a distance-dependent bias. Specifically, the bias associated with agent $i$ and option $j$ is
\begin{equation}
    b_{ij} = -\kappa\, \lVert \mathbf{X}_i - \mathbf{X}_{Gj} \rVert .
    \label{eq:bias}
\end{equation}
where $\kappa >0$ is the bias adaptation rate.
Hence, an option becomes increasingly unfavorable in the opinion dynamics as the agent moves farther from its corresponding sampling site.
This bias provides a feedback loop between the agents' physical position and the corresponding opinion vector: opinions determine where the agents move, while the resulting position modifies the bias of the opinion evolution.

The position dynamics of the agents are as follows:
\begin{equation}
    \dot{\mathbf{X}}_i = -k \sum_{j=1}^{\No} w_{ij}\,(\mathbf{X}_i - \mathbf{X}_{Gj}),
    \qquad \mathbf{w}_i = \sigma_{\mathcal{A}_i}(\beta_x \mathbf{Z}_{i}),
    \label{eq:position}
\end{equation}
with saturated speed $\lVert \dot{\mathbf{X}}_i \rVert \le v_{\max}$, spatial control gain $k>0$, and masked softmax defined in Eqn.~\eqref{eq:masked_softmax_definition}.
Here, $\beta_x>0$ determines how strongly the position controller responds to differences in opinion. Each agent moves toward a weighted combination of the sampling sites, with the weights determined by its current opinion vector.

Consider the nonlinear saturation function acting on the whole opinion vector as a shifted masked softmax. 
\begin{equation}
    S_i(\mathbf{v}) = \sigma_{\mathcal{A}_i}(\beta_s \mathbf{v}) - \tfrac{1}{|\mathcal{A}_i|}
    \mathbf{1}_{\mathcal{A}_i},
    \label{eq:wta}
\end{equation}
which is winner-take-all across options and sums to zero over
$\mathcal{A}_i$. 
Here, $\beta_s>0$ controls the sensitivity of the saturation, $\mathbf{1}_{\mathcal{A}_i}$ is the indicator vector of the capability set, and $|\mathcal{A}_i|$ gives the cardinality of the capability set.  This construction has two important properties. First, the saturation assigns zero weight to options outside $\mathcal{A}_i$. Second, its components sum to zero over the available options, i.e., $\sum_{j\in\mathcal{A}_i}[S_i(\mathbf{v})]_j=0.$

Consequently, the opinion dynamics evolves in the capability subspace
\begin{equation}
V_{i} = \{ \mathbf{Z}_{i} : z_{ij} = 0 \ \forall j \notin \mathcal{A}_i,\ \sum_{j} z_{ij} = 0 \}
\end{equation}
which is invariant under the opinion flow. 
Opinions evolve in the Nonlinear Opinion Dynamics (NOD) form with the position dependence carried entirely
by the bias \eqref{eq:bias}:
\begin{equation}
    \tau_z \dot{\mathbf{Z}}_i =  -d\, \mathbf{Z}_{i}
    + u\, S_i\Big( \alpha \mathbf{Z}_{i}
    + \gamma \sum_{k \neq i} (A_a)_{ik} \mathbf{Z}_{k} + \mathbf{b}_{i} \Big).
    \label{eq:F1}
\end{equation}
where $\mathbf{b}_{i}=(b_{i1},\ldots,b_{i\No})^\top$ and $i \in \{1,\cdots,\Na\}$.
Model parameters include the attention parameter ($u >0$), opinion damping coefficient ($d>0$), social influence weight ($\gamma \in \mathbb{R}$), and strength of self-reinforcement of opinion ($\alpha >0$). We assume that opinions evolve on a faster timescale than the physical movement of the agents, which is captured by the opinion timescale constant $\tau_z \ll 1$.

The social influence weight $\gamma$ serves as the behavioral control parameter in the opinion dynamics model. When $\gamma <0$, agents are antagonistic towards their neighbors and tend to suppress agreement with each other. We will show that for sufficiently negative values of $\gamma$, this leads the agents to disperse across distinct sampling sites, resulting in a distributed assignment. In contrast, when $\gamma >0$, agents are encouraged to align their opinions, promoting collective agreement. We will show that beyond a critical positive threshold, this alignment drives the agents toward consensus on a single sampling site, which corresponds to emergent rendezvous. 

Altogether, the combined agent dynamics are given by:
\begin{subequations}
\label{eq:agent_dynamics}
\begin{align}
\dot{\mathbf{X}}_i &= -k \sum_{j=1}^{\No} w_{ij}\,(\mathbf{X}_i - \mathbf{X}_{Gj}), \qquad \mathbf{w}_i = \sigma_{\mathcal{A}_i}(\beta_x \mathbf{Z}_{i}), \label{eq:agent_dynamics_x}\\
\tau_z \dot{\mathbf{Z}}_i &= -d\, \mathbf{Z}_{i}
    + u\, S_i\Big( \alpha \mathbf{Z}_{i}
    + \gamma \sum_{k \neq i} (A_a)_{ik} \mathbf{Z}_{k} + \mathbf{b}_{i} \Big) \label{eq:agent_dynamics_z}\\
b_{ij} &= -\kappa\, \lVert \mathbf{X}_i - \mathbf{X}_{Gj} \rVert \label{eq:agent_dynamics_bias}
\end{align}
\end{subequations}
where $i \in \{1,\cdots,\Na\}$, and $j \in \{1,\cdots,\No\}$.

\subsubsection{Price-mediated task capacity feedback} 
The dynamics in Eqn.~\eqref{eq:agent_dynamics} coordinate agents purely through the social gain $\gamma$, but does not explicitly account for external mission-relevant information such as the number of agents that a sampling site can accommodate. 
To address this, we assign each sampling site $\mathbf{X}_{Gj}$ a capacity $c_j \ge 0$, representing the task demand. 
For example, in the motivating scenario described in Section~\ref{sec:Motivating Example}, USVs can use preliminary search results to estimate the number of ASVs needed at each target site. They can then communicate these estimates to the AUVs to guide their allocation across the available sites. Once sufficient samples have been collected at a site, its capacity can be set to $c_{j}=0$, which discourages further agent assignments to that site.

Let \emph{price} $\lambda_j \ge 0$, be a dynamic penalty for committing to an over-subscribed site. These prices can be updated by the AUVs using the capacity information communicated by the USVs. 
We let the prices be dynamic quantities that evolve according to 
\begin{equation}
    \dot{\lambda}_j = \varepsilon \big( \ell_j(Z) - c_j \big), 
   \qquad
    \ell_j(Z) = \sum_{i=1}^{\Na}
    \big[\sigma_{\mathcal{A}_i}(\beta_\ell \mathbf{Z}_{i})\big]_j,
    \label{eq:price}
\end{equation}
where $\ell_j(Z)$ is the \emph{fractional load} on site $j$: each agent contributes its current soft commitment, computed as a masked softmax of its opinions with sharpness $\beta_\ell > 0$, so $\ell_j$ smoothly counts the agents presently favoring site $j$, and $\varepsilon > 0$ is the price adaptation rate. 

According to the dynamics in \eqref{eq:price} , the price $\lambda_j$ grows while site $j$ is loaded beyond its capacity, decays while it is underutilized, subject to the nonnegativity constraint on prices. Prices enter each agent's decision through a modification of the bias term $\mathbf{b}_i$ in the opinion dynamics, replacing \eqref{eq:agent_dynamics_bias} with 
\begin{equation}
\label{eq:bias_with_capacity}
    b_{ij} = -\kappa\, \lVert \mathbf{X}_i - \mathbf{X}_{Gj} \rVert - \lambda_j ,
\end{equation}
Consequently, an agent perceives an over-subscribed site as effectively more distant. Equations \eqref{eq:F1}, \eqref{eq:price} and \eqref{eq:bias_with_capacity} can be interpreted as a continuous-time analogue of an auction \cite{bertsekas1988auction}.
In the motivating scenario, sites that attract more AUVs than their estimated capacities become progressively more expensive, encouraging the AUVs to redistribute themselves towards other sites. As sampling progresses, the USVs can update the capacity information communicated to the AUVs, allowing completed sites to be assigned zero capacity, discouraging further assignments to them.
We assume that opinion evolution, spatial movement, and price updates evolve on separate timescales ordered as $\tau_z \ll 1 \ll \varepsilon^{-1}$. This way, opinions evolve the fastest, followed by the motion of physical trajectories to the target site locations, then followed by the adaptation in price associated to each site. This allows prices to avoid overreaction to transient fluctuations in agent opinions.

In our framework, social influence and price play complementary roles in coordinating the agents. 
The social gain $\gamma$ determines the amount of social influence exerted by agents on one another through communication of opinions between neighbors. On the other hand, the dynamic prices incorporate site capacities into this allocation process by adaptively tuning the bias in the opinion dynamics, in competition with the social effects and the proximity effects in the dynamic decision process.
Together, these mechanisms can improve the distribution of agents across sampling sites relative to coordination based on social influence alone, while also accounting for the differences in sampling demand over time and the completion of sampling tasks.

\subsection{Analysis of softmax nonlinear opinion dynamics for a single agent}
The NOD framework uses a component-wise bounded saturation function such as $\tanh$ to encode nonlinearity in the opinion dynamics \cite{bizyaeva2023tac,bizyaeva2025tac,leonard2024fast}. This framework works well for task allocation problems involving up to two options \cite{bizyaeva2022switching}. However, when more than two options are considered,  the $\tanh$ nonlinearity can give equilibria for our application purposes, such as ones where multiple tasks simultaneously have large associated opinion values at equilibrium. This can cause a moving agent to stop between multiple sites instead of converging to a specific sample site. To address this issue, we used the shifted masked softmax nonlinearity given in Eqn.~\eqref{eq:wta}. This choice is inspired by the deterministic annealing methods for combinatorial optimization based on Potts neurons \cite{david2022deterministic} and by logit dynamics from evolutionary game theory \cite{sandholm2010population,cianfanelli2025stability}. The shifted masked softmax promotes winner-take-all behavior in the opinion dynamics, causing every agent to converge to a specific sample site. To illustrate the decision-making bifurcation mechanism in this model, in the remainder of this section we will sketch out the bifurcation analysis for the simple case of a single agent's decision dynamics.

For an agent following Eqn.~\eqref{eq:agent_dynamics_z}, the opinion dynamics in the absence of social influence and with a static bias is given by
\begin{equation}
\label{eq:opinion_dyanmics_1agent_2option}
\tau_z \dot{\mathbf{Z}}_1 =  -d\, \mathbf{Z}_1 + u\, S_1\Big( \alpha \mathbf{Z}_1 + \mathbf{b}_1 \Big)    
\end{equation}
Let $\mathcal{A}_{1} = \{1,\cdots,\No\}$ be the capability set of the agent. 
For $\mathbf{b}_{1} = \mathbf{0}_{\Na}$, the undecided neutral state $\mathbf{Z}_{1}^{*} = \mathbf{0}_{\No}$ is always an equilibrium point of \eqref{eq:opinion_dyanmics_1agent_2option} and we study its local stability. The Jacobian evaluated at $\mathbf{Z}_{1}^{*}$, with $\mathbf{b}_{1} = \mathbf{0}_{\Na}$ is 
\begin{equation*}
J = \frac{1}{\tau_{z}}\left[-dI + \frac{u \alpha \beta_{s}}{\No} \left(I - \frac{1}{\No}\mathbf{1}_{\No}\mathbf{1}_{\No}^{\top} \right) \right]
\end{equation*}
where $I$ is the identity matrix, $\mathbf{0}_{\No}$ and $\mathbf{1}_{\No}$  denote the vectors of all zeros and ones, respectively. The eigenvalues of the Jacobian $J$ are $\lambda_{0} = -\frac{d}{\tau_{z}}$ with multiplicity 1 and $\lambda_{1} = -d + \frac{u \alpha \beta_{s}}{\No}$ with multiplicity $\No - 1$. The eigenvector associated with $\lambda_{0}$ is $\mathbf{1}_{\No}$.
The critical value of attention at which the neutral equilibrium changes stability is $u^{*} = \frac{d \No}{\alpha \beta_{s}}$.
Hence, $\mathbf{Z}_{1}^{*} = \mathbf{0}_{\No}$ is locally asymptotically stable within the zero-sum subspace for $u<u^{*}$ and loses stability for $u > u^{*}$. At $u=u^{*}$, all $\No - 1$ eigenvalues simultaneously cross zero, which gives rise to a bifurcation in the opinion dynamics.

For $\No \geq 2$ options or tasks, the zero-sum subspace of the opinion space $\mathbb{R}^{\No}$ has dimension $\No - 1$, and $\No-1$ eigenvalues of the Jacobian simultaneously become zero at the critical point. 
Furthermore, if the agent is unbiased, if the admissible set $\mathcal{A}_{1}$ contains all $\No$ options, i.e. $\mathcal{A}_{1} = \{1, \dots, \No\}$,  the dynamics \eqref{eq:opinion_dyanmics_1agent_2option} are $S_{\No}$-equivariant, i.e. equivariant under permutations of the options which means $F(P\mathbf{Z}) = P F(\mathbf{Z})$ where $P$ is a permutation matrix and $F: \mathbb{R}^{\No} \to \mathbb{R}^{\No}$ is the map defining right hand side of \eqref{eq:opinion_dyanmics_1agent_2option} \cite{golubitsky2003symmetry}. 
The bifurcation in the opinion dynamics model when indecision loses stability then corresponds to the $S_{\No}$-equivariant bifurcation discussed in \cite{elmhirst2004s}, analogous to the bifurcation analysis for all-to-all coupled NOD in the original saturating function formulation \cite{franci2023breaking}. 

We can use the $S_{\No}$-equivariance of the dynamics \eqref{eq:opinion_dyanmics_1agent_2option} to classify its bifurcation diagram by studying a one-dimensional reduced system on an appropriate fixed-point subspace. The Equivariant Branching Lemma \cite[Theorem 3.3]{golubitsky2012singularities} guarantees the existence of a branch of solutions, called the primary branch. 
Up to permutation, the primary-branch equilibria lie in one-dimensional subspaces parameterized by
\begin{equation*}
    \mathbf{Z}_1 = a\,\mathbf{v}_{p,q},
\qquad
\mathbf{v}_{p,q}
=
\bigl(
\underbrace{q,\ldots,q}_{p},
\underbrace{-p,\ldots,-p}_{q}
\bigr)^\top,
\qquad a\in\mathbb{R}.
\end{equation*}
where $p \leq q$, and $p + q = \No$ \cite{elmhirst2004s}.
We are particularly interested in the case $p = 1$ and $q = \No-1$, which corresponds to winner-take-all decision behavior in which one opinion dominates over all others whenever $a > 0$. It turns out that the bifurcation captured by the softmax opinion dynamics \eqref{eq:opinion_dyanmics_1agent_2option} is indeed the winner-take-all bifurcation with $p = 1$.

To understand the emergence of the winner-take-all decision behavior, consider the unbiased agent with $\mathbf{b}_{1} = \mathbf{0}_{\No}$. At an equilibrium satisfying $\dot{\mathbf{Z}}_1 = \mathbf{0}_{\No}$, define $w_{j} = \frac{d}{u}z_{1j} + \frac{1}{\No}, \forall j \in \{1,\cdots,\No\}$. Since, $\sum_{j=1}^{\No} z_{1j} = 0$, the probability vector $\mathbf{w} = (w_{1},\cdots,w{\No})^{\top} \in \mathbb{R}^{\No}$ satisfies $\sum_{j=1}^{\No}w_{j} = 1$. The equilibrium condition yields the following self-consistency equation:
\begin{equation*}
w_{j} = \frac{\exp (u\alpha \beta_{s} w_{j} / d)}{\sum_{k=1}^{\No} \exp (u\alpha \beta_{s} w_{k} / d)}
\end{equation*}
This equilibrium condition coincides with the mean-field self-consistency equation studied in statistical mechanics for the
Curie-Weiss-Potts (CWP) model with $\No$ states and zero external field \cite{gandolfo2010potts}, under the mapping of the inverse temperature parameter from the statistical physics model to the opinion dynamics parameters as $\beta = u\alpha \beta_{s} / d$.
This correspondence allows us to use established results for the CWP model to characterize equilibria of the opinion dynamics.
Hence, for fixed $\alpha, \beta_{s}$ and $d$, increasing the bifurcation parameter $u$ corresponds to increasing the inverse-temperature parameter of the CWP model. 
Let $\beta_{c}$ denote the critical inverse temperature where the phase transition occurs for the CWP model. 
For $\No>2$, \cite[Theorem 2.3]{gandolfo2010potts} shows that, when $\beta < \beta_{c}$, the free energy has a unique global minimizer at the uniform distribution $\mathbf{w} = (\frac{1}{\No},\cdots,\frac{1}{\No})$, which corresponds to the neutral equilibrium $\mathbf{Z}_{1}^{*} = \mathbf{0}_{\No}$ being stable for $u<u_{1}<u^{*}$, where $u_{1} = \frac{d \beta_{c}}{\alpha \beta_{s}}$. 
For $\beta>\beta_c$, equivalently $u>u_1$, the free energy has $\No$ symmetry-related global minimizers, corresponding to winner-take-all equilibria on the $p=1$ primary branches in the opinion dynamics, with a different option uniquely favored while the remaining $N_o-1$ are discounted at each equilibrium. We will next illustrate the symmetric bifurcation explicitly for 2 and 3 options, and numerically illustrate the effect of biases on these solution branches.

\subsubsection{$\No = 2$}
\begin{figure}[!htbp]
    \centering
    \begin{subfigure}{0.49\linewidth}
    \centering
    \includegraphics[width=\linewidth]{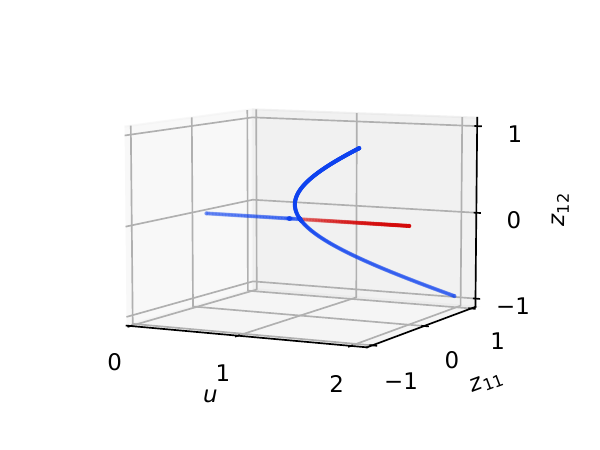}
    \caption{$\No = 2$, $\mathbf{b}_{1} = \left(0,0\right)$}
    \label{fig:1_agent_y_vs_u_2options}
    \end{subfigure}
    \hfill
    \begin{subfigure}{0.49\linewidth}
    \centering
    \includegraphics[width=\linewidth]{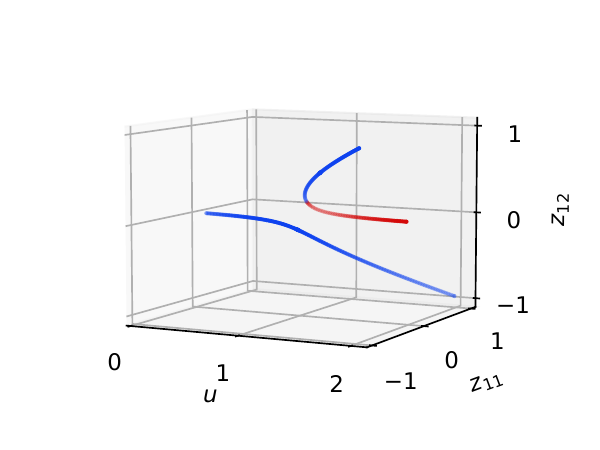}
    \caption{$\No = 2$, $\mathbf{b}_{1} = \left(0.05,0\right)$}
    \label{fig:1_agent_y_vs_u_2options_biased}
    \end{subfigure}
    \\
    \begin{subfigure}{0.49\linewidth}
    \centering
    \includegraphics[width=\linewidth]{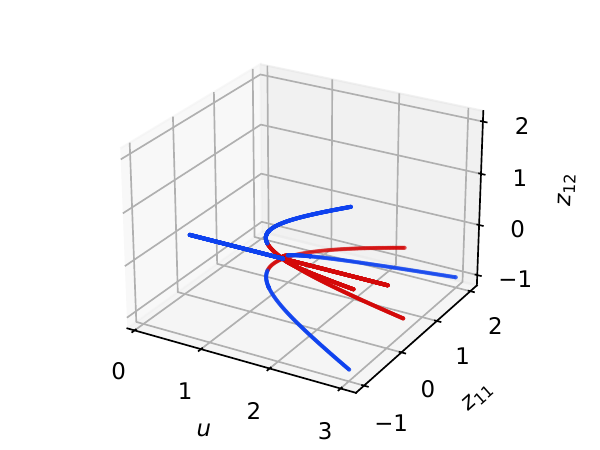}
    \caption{$\No = 3$, $\mathbf{b}_{1} = \left(0,0,0\right)$}
    \label{fig:1_agent_y_vs_u}
    \end{subfigure}
    \hfill
    \begin{subfigure}{0.49\linewidth}
    \centering
    \includegraphics[width=\linewidth]{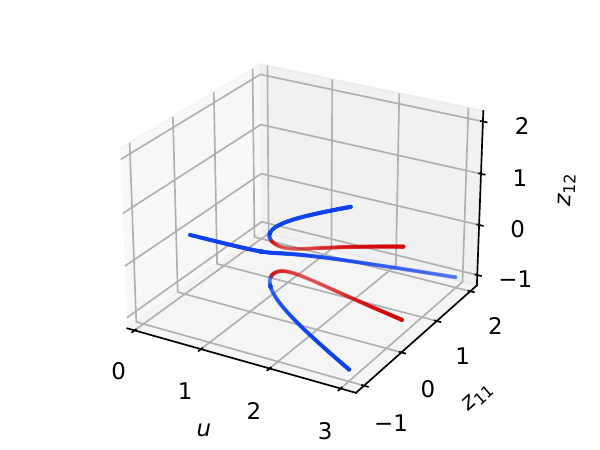}
    \caption{$\No = 3$, $\mathbf{b}_{1} = \left(0.05,0,0\right)$}
    \label{fig:1_agent_y_vs_u_biased}
    \end{subfigure}
    \caption{One agent, two/three option bifurcation diagrams with parameters $\alpha = 1.0$, $d = 1.0$, $\beta_{s} = 2.0$, and $\tau_{z} = 0.15$. 
    Blue (red) curves denote stable (unstable) equilibria.
    (a) and (b) show three-dimensional bifurcation diagrams in which the equilibrium opinions of the agents' two options are plotted for varying $u$. 
    For the unbiased case (a), at $u^{*} = 1.0$, two stable equilibrium branches emerge, each corresponding to the agent selecting one of the two distinct options (tasks).
    In the biased case (b), the unfolding favors option 1.
    (c) and (d) show three-dimensional bifurcation diagrams in which the equilibrium opinions of the first two options are plotted for varying $u$. 
    For the unbiased case (c), with sufficiently large $u$, three stable equilibrium branches emerge, each corresponding to the agent selecting one of the three distinct options (tasks). 
    In the biased case (d), the unfolding favors option 1. Numerical bifurcation diagrams were computed with \texttt{BifurcationKit.jl} \cite{veltz:hal-02902346}.
    }
\end{figure}

For two options, the opinion dynamics in Eqn.~\eqref{eq:opinion_dyanmics_1agent_2option} reduce to the standard $\tanh$ nonlinearity. Let $\mathbf{Z}_1 = (z_{11},z_{12})^{\top} \in \mathbb{R}^{2}$ be the opinion and $\mathcal{A}_1 = \{1,2\}$ be the capability set of the agent. Since the zero-sum structure of the opinions is preserved, the opinion dynamics can be written as:
\begin{equation*}
\tau_z \dot{z}_{11} =  -d\, z_{11} + \frac{1}{2}u\, \tanh \Big( \alpha \beta_{s} z_{11} + \frac{1}{2}\left(b_{11}-b_{12}\right) \Big); \quad z_{12} = -z_{11}
\end{equation*}
The above system recovers exactly the $\tanh$ formulation of the opinion dynamics in the original NOD framework \cite{bizyaeva2023tac}. The hyperbolic tangent nonlinearity naturally captures the symmetry for a binary decision, and the shifted softmax employed in our model formulation in this chapter naturally generalizes this to the setting of multiple mutually exclusive options. The bifurcation diagrams of the binary choice opinion equation are shown in Fig.~\ref{fig:1_agent_y_vs_u_2options} and Fig.~\ref{fig:1_agent_y_vs_u_2options_biased}. The unbiased case in Fig.~\ref{fig:1_agent_y_vs_u_2options} is the standard pitchfork bifurcation in which the neutral equilibrium loses stability, and two simultaneously stable branches of equilibria emerge, with each representing one of the two choices the agent can take on. When an agent has a nonzero bias, the symmetric pitchfork bifurcation diagram unfolds into a continuous solution branch and a saddle-node bifurcation, with the bias selecting the preferred option near the bifurcation point as illustrated in Fig.~\ref{fig:1_agent_y_vs_u_2options_biased}.

\subsubsection{$\No = 3$}
Let $\mathbf{Z}_1 = (z_{11},z_{12},z_{13})^{\top} \in \mathbb{R}^{3}$ be the opinion and $\mathcal{A}_1 = \{1,2,3\}$ be the capability set of an agent following the opinion dynamics in Eqn.~\eqref{eq:opinion_dyanmics_1agent_2option}. The numerically computed bifurcation diagrams of this agent are shown in Fig.~\ref{fig:1_agent_y_vs_u} and Fig.~\ref{fig:1_agent_y_vs_u_biased}. For the unbiased case in Fig.~\ref{fig:1_agent_y_vs_u}, the neutral state loses stability and three simultaneously stable equilibria emerge, each representing commitment to a single option.
This illustrates how winner-take-all choice emerges through a bifurcation in the model for $\No > 2$ options. For larger option sets, an analogous bifurcation is predicted, with the total number of stable branches equal to the option set dimension. When a bias for one of the options is introduced, Fig.~\ref{fig:1_agent_y_vs_u} illustrates the unfolding of the multidimensional bifurcation diagram. Analogously to the two-option case, near a bifurcation point, the bifurcation diagram breaks up into a continuous branch of locally stable equilibria and several saddle-node bifurcations. The continuous branch is selected by the agent's bias, and the competing equilibria are selectively removed near the bifurcation point of the symmetric point, illustrating the sensitivity of the decision. We expect this effect to generalize to larger option dimensions; however, we note that the unfolding theory for such high-dimensional bifurcation has not been rigorously worked out; therefore, the unfolding behavior we are describing is a numerical observation that we conjecture should hold up generically. The bifurcation diagram and its unfolding also illustrates the effect of social influence, geometry, and task capacity on the agent's opinion dynamics for the task allocation problem we are considering. In the absence of these factors, the decision dynamics are organized through an $S_{\No}$-equivariant bifurcation and will cause an agent to choose a task to do at random, following its initial conditions. 
Adding social influence, geometric effects, and prices, effectively tunes the agent's bias and dynamically unfolds its symmetric bifurcation diagram, selecting a choice that has the most evidence for it as a result of the three combined effects.

\subsection{Connection to logit dynamics}
In large-scale multi-agent systems, evolutionary game theory \cite{sandholm2010population} studies the collective behavior that emerges from strategic interactions and payoff differences among agents. 
Such behavior can be described using evolutionary dynamics, which model how the distribution of actions evolves in response to reward differences.
An important framework in this context is population games, in which a continuum of agents is partitioned into a finite number of populations. 
Each population $i$ has population size $v_{i} \in \mathbb{R}^{\geq 0}$, action set $\mathcal{A}_{i}$, and a reward function that depends on the current population state. Agents belonging to the same population are indistinguishable and choose actions from their common action set $\mathcal{A}_{i}$. 

Let $Y = \{\mathbf{Y}_{1},\cdots,\mathbf{Y}_{\Na}\}$ denote the global population state, where $\mathbf{Y}_{i} = \left(y_{i1},\cdots,y_{i\No}\right) \in \mathbb{R}^{\No}$ describes the distribution of population $i$ across the available actions. In particular, $y_{ij} \geq 0$ denotes the mass of agents in population $i$ choosing action $j$, and the population state satisfies $\sum_{j=1}^{\No} y_{ij} = v_i$ with $y_{ij} = 0$ if $j \notin \mathcal{A}_{i}$.
Let $r_{ij}(Y)$ denote the reward associated with action $j$ in population $i$ at global population state $Y$. 
For a given population game, Nash equilibrium is defined as a state where every action played by a positive mass $y_{ij}$ yields the maximum reward among the actions available to that population.

A widely studied class of evolutionary dynamics for population games is the logit dynamics, which models the agents' tendency to select actions with higher rewards while allowing suboptimal actions. The noise parameter $\eta \in (0, \infty)$ controls the extent of these deviations.
The continuous-time logit dynamics is as follows:
\begin{equation*}
\dot{y}_{ij} = 
\begin{cases}
    \dfrac{v_{i} \exp (r_{ij} (Y)/\eta)}{\sum_{l \in \mathcal{A}_{i}} \exp (r_{il} (Y)/\eta)} - y_{ij},  & j \in \mathcal{A}_{i},\\[2mm]
    0, & j \notin \mathcal{A}_{i},
\end{cases}
\end{equation*}
These dynamics preserve the population size constraint $\sum_{j=1}^{\No}y_{ij} = v_{i}$.

As $\eta \rightarrow 0^{+}$, the logit dynamics favors actions which maximizes the rewards, whereas as $\eta \rightarrow \infty$, $\mathbf{Y}_{i}$ converges to the uniform distribution over the available actions.
In \cite{cianfanelli2025stability}, the authors show that every strict Nash equilibrium of the logit dynamics gives rise to a locally asymptotically stable equilibrium for sufficiently small noise. On the other hand, for sufficiently large noise, they establish the existence of a globally exponentially stable equilibrium. These results suggest the emergence of bifurcation for logit dynamics in the presence of multiple strict Nash equilibria, analogous to decision bifurcations in NOD, which the authors illustrate using numerical examples in \cite{cianfanelli2025stability}. 

The logit dynamics are closely related to the opinion dynamics studied in this chapter.
In particular, consider the choices, $v_{i} = \frac{u}{\tau_{z}}$, $\eta = \frac{1}{\beta_{s}}$, $\mathbf{Y}_{i} = \frac{d}{\tau_{z}}\mathbf{Z}_{i} + \frac{u}{\tau_{z}|\mathcal{A}_{i}|}\mathbf{1}_{\mathcal{A}_{i}}$, and rewards function vector $r_{i}(Z) = \alpha \mathbf{Z}_{i}+ \gamma \sum_{k \neq i} (A_a)_{ik} \mathbf{Z}_{k} + \mathbf{b}_{i}$.
Under this affine transformation, the logit dynamics is related to the opinion dynamics in Eqn.~\eqref{eq:agent_dynamics_z}, provided the admissible action sets agree with the capability sets used in the opinion model. 
This connection allows us to interpret the nonlinear opinion dynamics through the lens of evolutionary game theory. 
Under the outlined mapping, each physical agent in our framework corresponds to a population whose actions are its feasible tasks, and the transformed opinion vector describes the distribution of that population across actions. The self-reinforcement term increases the reward of an already favored task, while the social term rewards alignment with neighboring agents when $\gamma>0$ and discourages it when $\gamma<0$. These interactions therefore provide incentives for cohesive and dispersed allocations, respectively, which we will illustrate numerically in the following section.

\section{Behavioral regimes of the coupled task allocation dynamics}
\label{sec:Behavioral regimes of the coupled dynamics}
In this section, we illustrate via simulation studies representative collective behaviors that emerge from the dynamics discussed in Sec.~\ref{subsec:Agent dynamics}.
Consider a group of $\Na$ agents that must collectively select among $\No$ sample sites. Unless mentioned otherwise, the parameters used for the simulations in this section are $k = 0.8$, $\beta_x = 4.0$, $v_{\max} = 0.4$, $\beta_{s} = 2.0$, $\kappa = 2.0$, $u = 2.0$, $d = 1.0$, $\alpha =1.0$, and $\tau_{z} = 0.15$. 
For all the simulations discussed in this section, the initial opinion vector $Z_{i}(0)$ of each agent $i$ is initialized with small, zero-mean Gaussian random perturbations with a standard deviation of $10^{-3}$. These perturbations are masked by the agent's capability set $\mathcal{A}_{i}$ and then projected onto the zero-sum subspace. This initialization ensures that the initial opinions have small magnitudes, satisfy the zero-sum constraint, and are zero for all options outside the agent's capability set. 

\subsection{Effect of social coupling $\gamma$}

First, we investigate how the macroscopic coalition formation patterns are shaped by the agents' social relationships under all-to-all communication.  The social coupling parameter $\gamma$ determines whether agents are cooperative, i.e., reinforce each others' opinions,  or antagonistic, i.e., suppress each others' opinions.
We expect the opinion dynamics to promote consensus choice on a task when $\gamma>0$ and dispersed allocations when $\gamma<0$, which we will illustrate in simulation. Our simulations illustrate how by varying just this single parameter, a large set of collective allocation patterns can be reached from identical initial conditions, with predictable trends in group cohesion as agents are made more cooperative or more antagonistic.

Consider $\Na =10$ agents and $\No = 10$ sample sites, with the agents following the dynamics given in Eqn.~\eqref{eq:agent_dynamics} and interacting over a complete graph. 
Fig.~\ref{fig:F1_9} illustrates how increasing $\gamma$ from a sufficiently negative value to a sufficiently positive value tunes group cohesion through a sequence of macroscopic states, from a maximally dispersed allocation with one agent at each sampling site, through intermediate allocations with several self-organized subgroups, to consensus with all agents gathered at a single site.

This tunable group cohesion observed in simulation illustrates how the same set of decision rules can support different task allocation objectives. 
In particular, for different choices of $\gamma$, the resulting allocation pattern can be interpreted within the taxonomy of multi-robot task allocation in \cite{gerkey2004formal}. If the number of distinct clusters is less than $\No$, then some of the sample sites receive multiple agents, which corresponds to the ``single-task robots, multi-robot tasks, instantaneous assignment'' (ST-MR-IA) problem, also known as coalition formation \cite{gerkey2004formal}. The equivalent optimization problem is the set partitioning problem, i.e., partition the $\Na$ agents among $\No$ sample sites, where each sample site can be assigned multiple agents.

In contrast, if the number of distinct clusters is equal to $\No$, then the agents are maximally dispersed, which corresponds to the ``single-task robots, single-robot tasks, instantaneous assignment'' (ST-SR-IA) problem \cite{gerkey2004formal}. This is equivalent to the optimal assignment problem, i.e., we want to distribute $\Na$ agents among $\No$ (=$\Na$) options, where each sample site is assigned exactly one agent.
Next, we will investigate the quality of the resulting assignments in representative parameter regimes.

\subsection{Dispersion and optimality gap from the Hungarian optimum for maximally antagonistic agents}

First we consider a group of antagonistic agents with $\gamma < 0$, assuming the social coupling is sufficiently strong in magnitude to yield the maximally dispersed behavior observed in the previous subsection and shown in Fig.~\ref{fig:F1_9}.
We are interested in quantifying the quality of the selected allocation.  A natural benchmark to compare against is the Hungarian algorithm \cite{kuhn1955hungarian} for the optimal assignment problem. Let $P \in S_{\Na}$ be a permutation on $\Na$ symbols that represents the task assignment of agents to task sites, where $P(i)$ is the site assignment for agent $i$. The cost of the assignment for the permutation $P$ is defined in terms of the Euclidean distances between agents' initial positions and their assigned goal locations, 
\begin{equation}
    J(P) = \sum_{i=1}^{\Na} \| \mathbf{X}_i(0) - \mathbf{X}_{G_{P(i)}} \|_2.
\end{equation}
The Hungarian optimum is the permutation corresponding to the minimal cost $J_H = \min_{P\in S_{\Na}} J(P)$. Let $P_{NOD}$ be the task assignment achieved via the opinion-guided embodied strategy \eqref{eq:agent_dynamics} and define its corresponding assignment cost as $J_{NOD} = J(P_{NOD})$. In simulations summarized in Fig.~\ref{fig:F1_hist_complete_network_1} we evaluate the resulting assignment in terms of its optimality gap against the Hungarian optimum, defined as $(J_{NOD} - J_H)/J_H \times 100\%$.
 
Fig.~\ref{fig:F1_hist_complete_network_1} shows histograms of the optimality gap for initial agent positions randomly sampled from a uniform probability density function over the domain $\left[-0.9,0.9\right] \times \left[-0.9,0.9\right]$. As the number of sample sites increases, while keeping its domain constant, the average optimality gap tends to increases with the increased density of sample sites. However, even in these cluttered environments, most simulations result in optimal or close-to-optimal assignment.
Rigorously investigating conditions under which these dynamics guarantee optimal or near-optimal allocation and quantifying the optimality gap is an interesting direction for future work.

\subsection{Convergence to a single sample site for maximally cooperative agents}
\begin{figure}[ht]
    \centering
    \includegraphics[width=\linewidth]{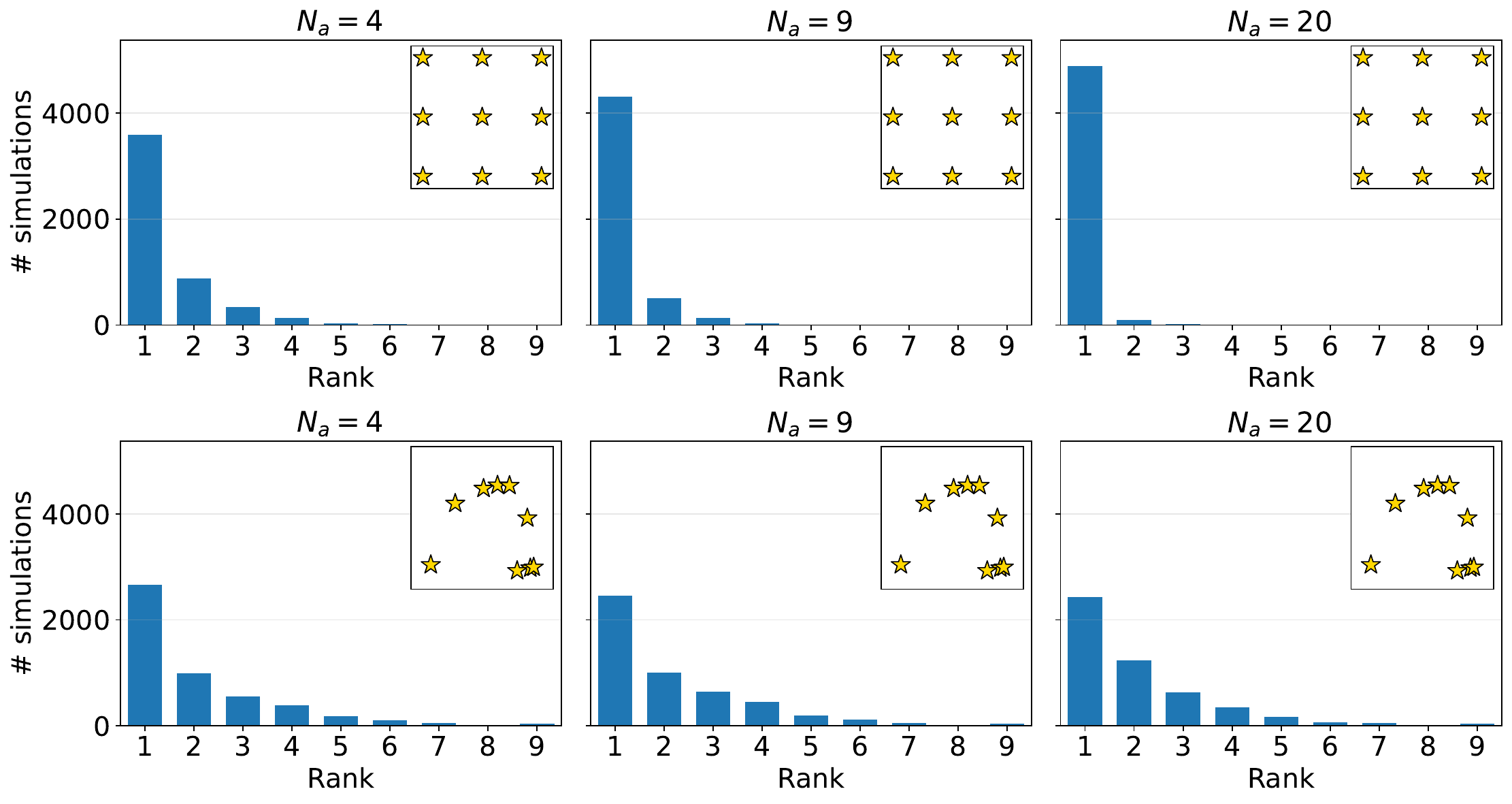}
    \caption{Consensus site selection for $\gamma=2$ and $\No=9$ sampling sites with all-to-all communication. Each histogram shows the rank of the site selected by the group over 5,000 simulations, with initial agent positions sampled uniformly at random from $[-0.9,0.9]\times[-0.9,0.9]$. Sites are ranked by increasing distance from the centroid of the agents' initial positions, with rank 1 denoting the nearest site. Columns correspond to $N_a=4,9,20$ agents; rows correspond to the two site configurations shown in the insets, where yellow stars mark sampling sites in $[-1,1]\times[-1,1]$. The nearest site is the most frequently selected in all cases, while the distribution of selected ranks depends on the site geometry and group size.}
    \label{fig:F1_nearest_goal_rank_complete_network}
\end{figure}

Next we consider a group of cooperative agents with $\gamma > 0$, assuming the social coupling is sufficiently strong in magnitude to yield the consensus behavior shown in Fig.~\ref{fig:F1_nearest_goal_rank_complete_network} where all agents converge to a single site. 
Just as in the maximally dispersed case, we are interested in quantifying the quality of the consensus decision that emerges through the opinion and motion coupling. 
A natural reference for the consensus location is the point minimizing the sum of squared distances from the agents' initial positions, i.e., the centroid 
\begin{equation}
    \bar{\mathbf{X}}_0
    = \frac{1}{\Na}\sum_{i=1}^{\Na}\mathbf{X}_i(0).
\end{equation}
The centroid of initial conditions is the consensus value in the average consensus problem and provides a natural reference for rendezvous, as it minimizes the sum of squared distances from those positions.
Therefore, for best performance, we expect the agents to select a task site closest to their initial centroid $\bar{\mathbf{X}}_0$.
Let the sample sites be ranked on their proximity to the $\bar{\mathbf{X}}_0$, i.e., in increasing order of the distance
$\|\mathbf{X}_{G_j}-\bar{\mathbf{X}}_0\|_2$, with rank 1 denoting the nearest site, rank 2 the second nearest, and so on.  Fig.~\ref{fig:F1_nearest_goal_rank_complete_network} shows histograms of the sample sites to which all the agents converge across representative simulations. As before, for the simulations shown, the initial positions of the agents are sampled from a uniform probability density function over the domain $\left[-0.9,0.9\right] \times \left[-0.9,0.9\right]$. 

Our simulation studies show that the site nearest to the initial centroid is the most frequently selected across different group sizes and site geometries, which suggests that the emergent consensus decision favors decisions that minimally displace the center of mass of the group within the environment. Across both regular grid environments and randomly placed target configurations, we see that decisions most frequently concentrate in the highest ranked target site, and the frequency of site selection falls off monotonically as a function of their distance to the initial centroid. These results motivate future investigation of the interaction between site geometry and initial agent placement in informing the ultimate site selection in the coupled opinion-motion dynamics.

\subsection{Intermediate clustered parameter regimes}

\begin{figure}[ht]
    \centering
    \includegraphics[width=\linewidth]{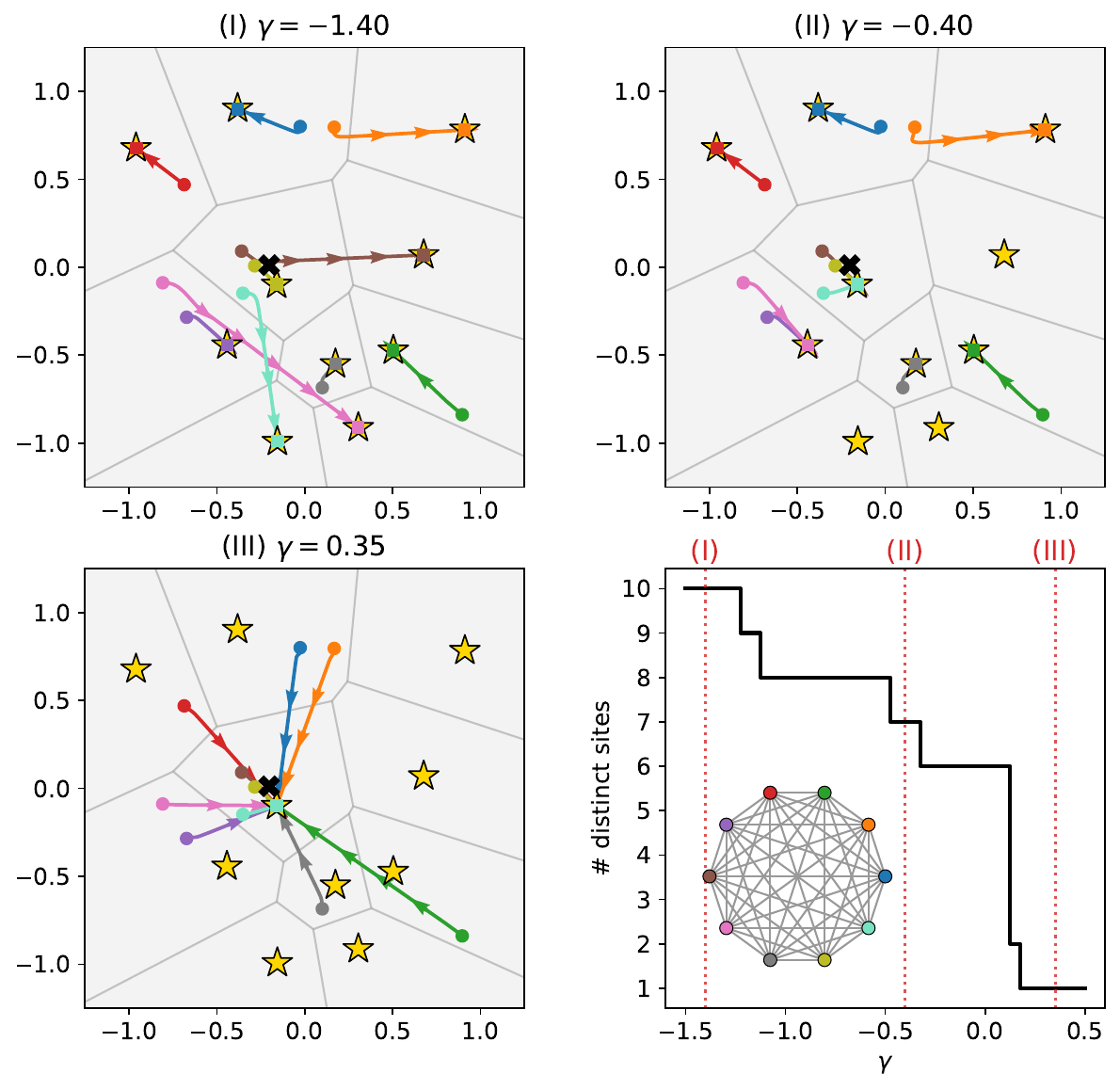}
    \caption{Simulation results for $\Na = 10$ agents and $\No = 10$ sample sites. The first three trajectory subplots show agent trajectories for three distinct values of $\gamma$, illustrating the formation of clusters of varying sizes.
    The bottom right subplot shows the number of distinct sample sites occupied by the agents as a function of $\gamma$. At each value of $\gamma$, simulated agents start from the same initial positions and sample site distribution. An inset shows the agents interacting over a complete graph, the communication topology used in these simulations.
    Yellow stars denote the $\No$ sample sites, while black crosses denote the centroid of the initial agent positions. Agent trajectories are shown in different colors, with circles and squares marking the initial and final positions, respectively.}
    \label{fig:F1_9}
\end{figure}

In Fig.\ref{fig:F1_9} we illustrate that as $\gamma$ is varied continuously from negative to positive, the emergent macroscopic task allocation pattern in the group transitions through multiple distinct regimes. As agents become more cooperative, they form a smaller number of total coalitions, ultimately resulting in a consensus formation when $\gamma$ is sufficiently large. This shows how a large number of operating regimes with distinct levels of cohesiveness in the group can be supported by the same decision mechanism. In future work,  we will explore how the geometry of the task environment and the agents' initial position shape these parameter regimes, and how groups can be controlled to flexibly transition between these different macroscopic allocation patterns through decentralized feedback on their social weights.

\subsection{Tunable cohesiveness with sparse communication}

\begin{figure}[ht]
    \centering
    \includegraphics[width=\linewidth]{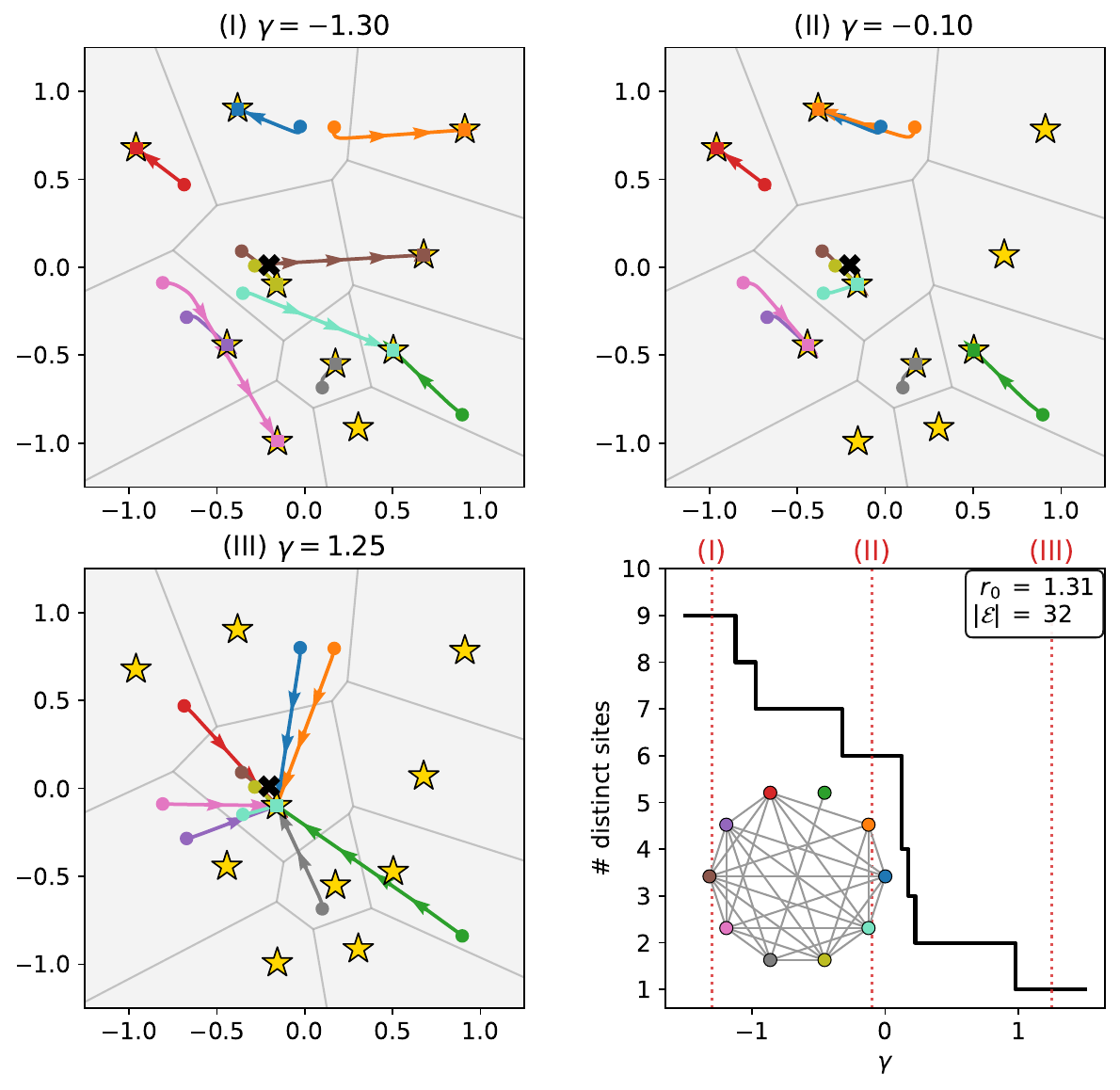}
    \caption{Simulation results for $\Na = 10$ agents and $\No = 10$ sample sites. The first three trajectory subplots show agent trajectories for three distinct values of $\gamma$, illustrating the formation of clusters of varying sizes.
    The bottom right subplot shows the number of distinct sample sites occupied by the agents as a function of $\gamma$. At each value of $\gamma$, simulated agents start from the same initial positions with the same task distribution. An inset shows the interaction graph among the agents that documents the communication topology used in the simulations.
    Yellow stars denote the $\No$ sample sites, while black crosses denote the centroid of the initial agent positions. Agent trajectories are shown in different colors, with circles and squares marking the initial and final positions, respectively.}
    \label{fig:F1_11}
\end{figure}

\begin{figure}[ht]
    \centering
    \includegraphics[width=\linewidth]{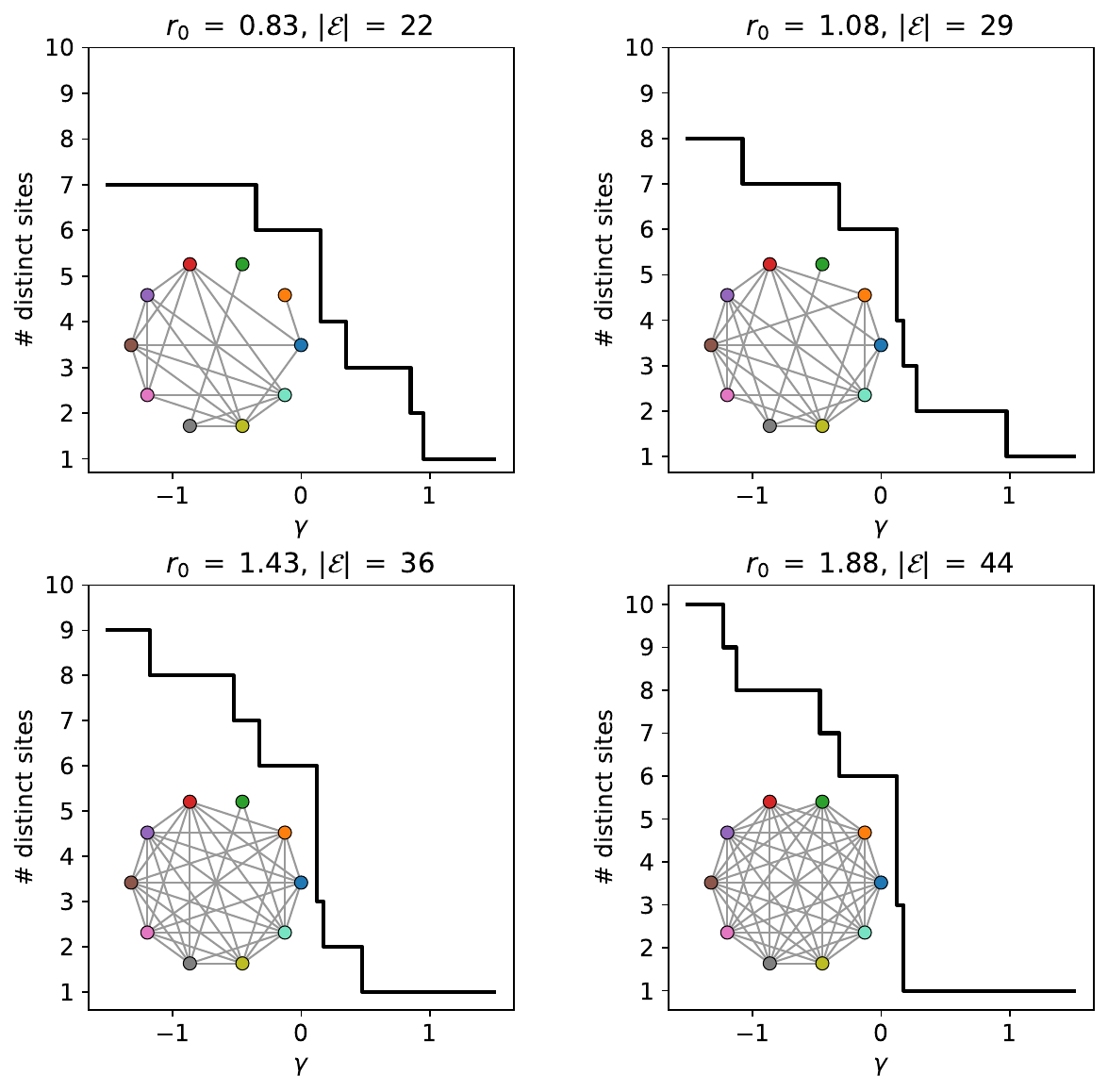}
    \caption{Parameter sweep simulation results for $\Na = 10$ agents and $\No = 10$ sample sites. At each value of $\gamma$, simulated agents start from the same initial positions and sample site distribution used in  Fig.~\ref{fig:F1_11}. The subplots show the number of distinct sample sites occupied by the agents as a function of $\gamma$ for a given $r_{0}$. The inset shows the interaction graph among the agents for the presented set of simulations.}
    \label{fig:F1_17}
\end{figure}

In the simulation results discussed so far, agents' communication was all-to-all. In this section, we relax this assumption to a slightly more realistic setting, letting the agents' communication networks be determined by their initial proximity to each other.
Consider $\Na =10$ agents and $\No = 10$ sample sites, with the agents following the dynamics given in Eqn.~\eqref{eq:agent_dynamics} and interacting according to the adjacency matrix $\Aa$,
\[ (\Aa)_{ii}=0, \qquad (\Aa)_{ij} = \begin{cases} 1, & \|\mathbf{X}_i(0)-\mathbf{X}_j(0)\| \le r_0,\\[4pt] 0, & \|\mathbf{X}_i(0)-\mathbf{X}_j(0)\| > r_0, \end{cases} \quad i\neq j. \] 
Note that the topology of the communication graph is set by the initial proximity of agents, but remains static for the duration of the simulation. 
Fig.~\ref{fig:F1_11} shows the transitions of the agents' final states as  $\gamma$ varies for a fixed $r_{0}$. As $\gamma$ increases, the agents' final states transition from multiple distinct clusters toward convergence to a single site. Here, for the given $r_{0}$, the interaction graph is not complete. Moreover, over the considered range of  $\gamma$, we do not observe the final states transitioning to $\Na = 10$ distinct clusters. This is because the chosen value of $r_{0}$ does not result in a sufficiently high number of interactions among the agents. Fig.~\ref{fig:F1_17} shows the effect of varying $r_{0}$ on the number of distinct sites, while keeping the sample sites and the initial agent states the same as in Fig.~\ref{fig:F1_11}. Interestingly, the final states transition to $\Na = 10$ distinct clusters even when the agent interaction graph is not complete.  These simulations suggest that tunable group cohesion observed in the all-to-all case persists under local communication, although the extent of dispersion achievable depends on the connectivity of the interaction network.

\subsection{Responsiveness to changes in site capacity}

Next, we consider the effects of the augmented bias \eqref{eq:bias_with_capacity} that encodes capacity limits of different target sites on the task selection in the group. The behaviors discussed until now have shown agent coordination purely through social influence modulated by the social gain $\gamma$.
In realistic scenarios, the capacity of the sample sites is often limited due to physical constraints or mission demands and can change during operation as information becomes available or the tasks are completed.
Consider the situation where we first send individual agents to distinct sample sites to obtain some preliminary data. Based on this data, the knowledge of the capacities of the sample sites can then be updated and synchronized within the team. This new information should in turn reorganize the macroscopic organization of the group across the sites to match the newly discovered demand.

\begin{figure}[ht]
    \centering
    \includegraphics[width=\linewidth]{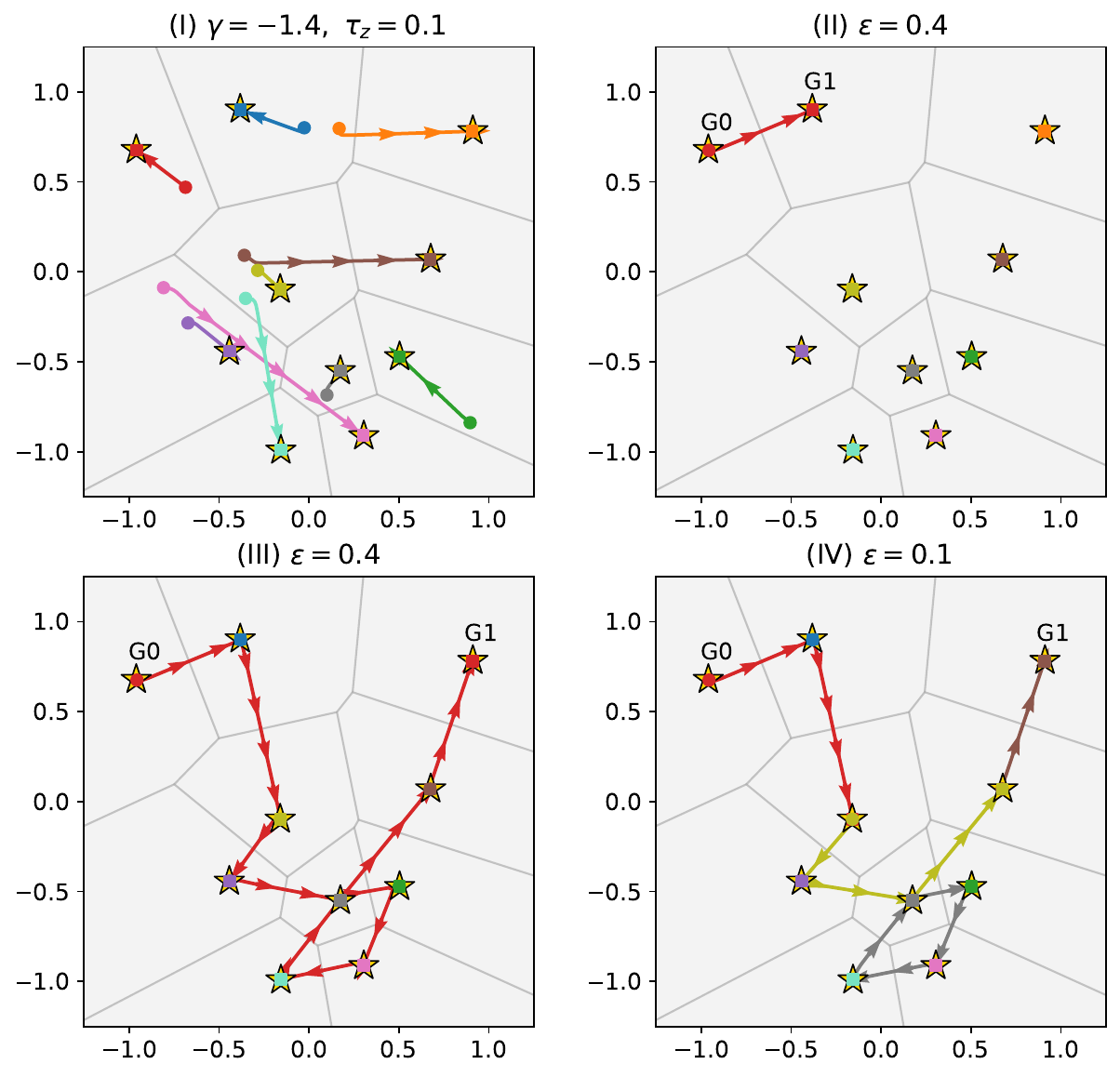}
    \caption{Simulation results for $\Na = 10$ agents interacting over a complete graph and $\No = 10$ sample sites, with $\tau_{z} = 0.1$, and $\gamma = -1.4$ in all the simulations. (I) shows convergence to $\Na$ distinct sample sites under the dynamics in Eqn.~\eqref{eq:agent_dynamics}. (II)$-$(IV) show the subsequent agent trajectories initialized at the final positions obtained in (I), under the capacity-dependent bias dynamics in Eqn.~\eqref{eq:bias_with_capacity}. The capacities are set such that site G0 is closed $(c_{j} = 0)$, site G1 can accommodate two agents $(c_{j} = 2)$, and all other sample sites can accommodate one agent each $(c_{j} = 1)$. In (II), sites G0 and G1 are close to each other, allowing the agent initially at G0 to move directly to G1. In (III) and (IV), these sites are farther apart, causing the agent to pass through multiple intermediate sample sites before reaching G1. Depending on $\epsilon$, the agent may switch its selected site at these intermediate locations, with multiple such switches occurring in (IV). Yellow stars denote the $\No$ sample sites. Agent trajectories are shown in different colors, with circles and squares marking the initial and final positions, respectively.}
    \label{fig:F1_16}
\end{figure}

We simulate one such scenario in Fig.~\ref{fig:F1_16}, where agents have all-to-all communication. In the simulated scenario, at first the capacity of the different sites is unknown, and agents disperse across sites following \eqref{eq:agent_dynamics} as is shown in panel (I). 
After this initial stage, the sample site labeled $G0$ is discovered to have zero capacity, and another sample site labeled $G1$ discovered to have capacity 2 at which point the price-dependent term in the bias \eqref{eq:bias_with_capacity} becomes nonzero. The group must then reallocate its members by removing the agent from $G_0$ to another location, and by increasing the number of agents at $G_1$ to two. We investigate the emergent reallocation behavior for different choices of $G_1$ and different rates of price adaptation.

In panel (II) of Fig.~\ref{fig:F1_16}, we show the intuitive outcome where the agent from $G_0$ relocates directly to $G_1$ when $G_1$ is sufficiently close to $G_0$ in the target geometry.
When the higher-capacity site $G_1$ is far away from the lower-capacity site $G_0$, we observe different reallocation behavior that is shaped in part by the price adaptation rate.  Interestingly, rather than inducing movement directly from $G_0$ to $G_1$, the emergent trajectories bounce from current target site to nearest unvisited target site until the allocation matches the site capacity distribution.
In one limit of this behavior, when the price adaptation rate $\epsilon$ is high, only one agent moves, as we show in panel (III). The agent that is initially assigned to site $G_0$ proceeds to navigate a zigzagging sequence of target sites until it reaches $G_1$. In another limit, when the price adaptation rate $\epsilon$ is low, an agent only travels a short distance before passing on the movement burden onto another team member. This behavior is pictured in panel (IV) where the red agent initially located in $G_0$ navigates a sequence of two target sites, at which point it remains still, and the agent previously located in that site navigates through a sequence of several locations and knocks the next occupant out of its original site. Scenarios (III) and (IV) result in the same total amount of movement, and in both cases the desired allocation distribution is reached, but in (III) the movement burden is taken on by a single unit, while in (IV) it is shared by the group in an emergent game of tag.

\begin{figure}[ht]
    \centering
    \includegraphics[width=\linewidth]{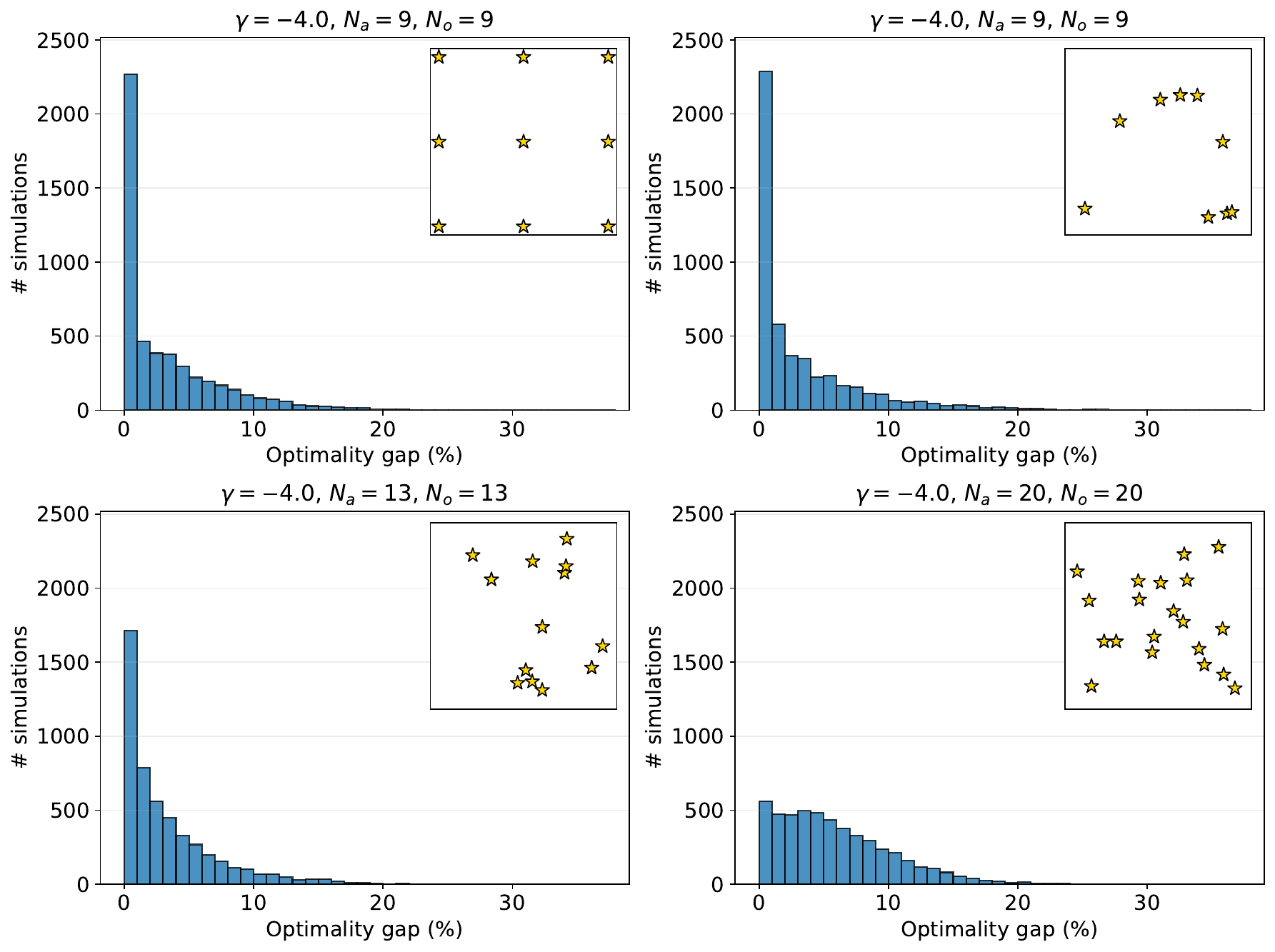}
    \caption{Numerical evaluation of optimality gap with all-to-all communication. Histograms show the percentage difference of the cost from the Hungarian optimal assignment cost obtained from 5,000 simulations from random initial conditions with indicated parameters and a fixed sample site distribution over the domain $\left[-1,1\right] \times \left[-1,1\right]$, pictured as an inset in each subplot. 
    }
    \label{fig:F1_hist_complete_network_1}
\end{figure}

\begin{figure}[ht]
    \centering
    \includegraphics[width=\linewidth]{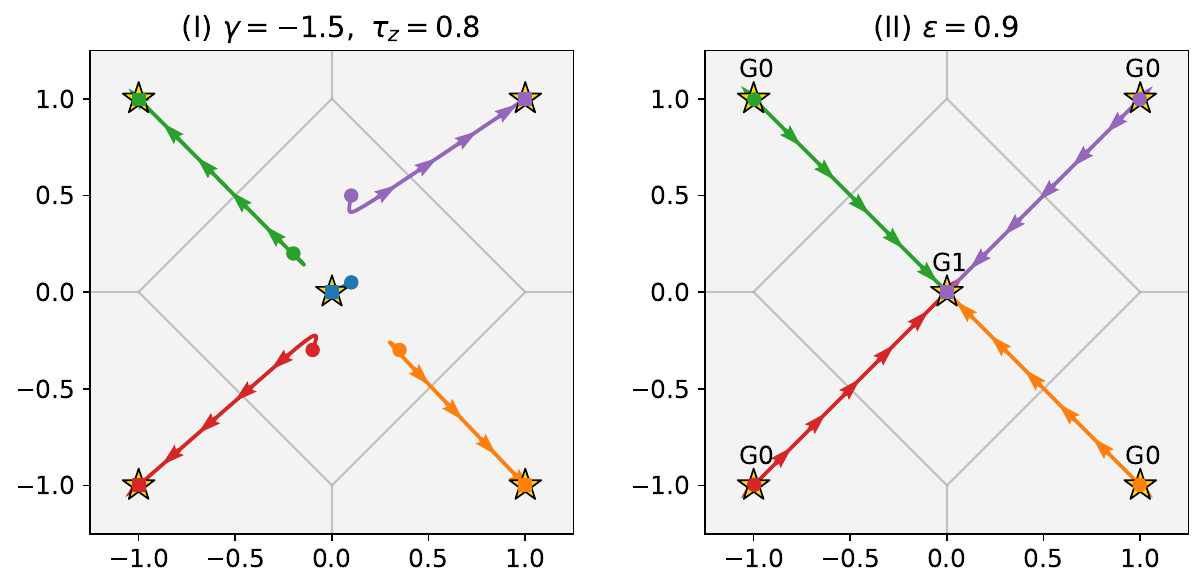}
    \caption{Simulation results for $\Na = 5$ agents interacting over a complete graph and $\No = 5$ sample sites, with $\tau_{z} = 0.8$, and $\gamma = -1.5$ in both simulations. (I) shows convergence to $\Na$ distinct sites under the dynamics in Eqn.~\eqref{eq:agent_dynamics}. (II) shows the subsequent agent trajectories, initialized at the final positions obtained in (I), under the capacity-dependent bias dynamics in Eqn.~\eqref{eq:bias_with_capacity}. The capacities are set such that sites marked $G0$ are closed $(c_{j} = 0)$, and site $G1$ can accommodate all $\Na$ agents $(c_{j} = \Na)$. In (II), the values of $\tau_{z}$, $\gamma$, and $\epsilon$ are chosen such that agents initially located at $G0$  move directly to $G1$. Yellow stars denote the $\No$ sample sites. Agent trajectories are shown in different colors, with circles and squares marking the initial and final positions, respectively.}
    \label{fig:F1_18}
\end{figure}

\section{AUV simulation study}
\label{sec:Simulations}
We evaluate the proposed distributed allocation approach for the motivating scenario in MOOS-IvP-Multi, a marine multi-agent autonomy framework extending the MOOS-IvP open-source autonomy project \cite{benjamin2010MOOSIvP}. Our goal is to examine whether the maximally dispersed allocation behavior observed above persists under communication and vehicle-motion constraints representative of a real-world AUV deployment, thus further validating the opinion-guided allocation strategy. The simulations focus on the distributed autonomy of the AUVs using the theoretical results developed above. A group of ten AUVs completes the distributed allocation process for ten sample site locations, each with a capacity of one, making the problem a perfect matching. The ten locations are randomly placed within a square 6km region.  Optimal placement of the USVs, which act as navigation beacons, is outside the scope of these simulations.

Communication range was limited to 3.5km, consistent with the typical ranges reported for the WHOI Micro-Modem \cite{Freitag2005TheWM} used in shelf and offshore environments. Each AUV broadcasts packets to all AUVs within range, with no acknowledgment or retransmission.  A linear sound-speed profile was assumed across the region. Packets were broadcast at 30 s intervals.
A dedicated IvP behavior implemented the motion update law, expressing preferences over heading and speed. Even with the long broadcast interval, the time-scale of the nonlinear opinion dynamics process was faster than the time-scale of the AUV motion, whose speed was capped at 1.0 m/s, consistent with the performance of low-cost vehicles.

\begin{figure}[ht]
\includegraphics[width=1.0\textwidth]{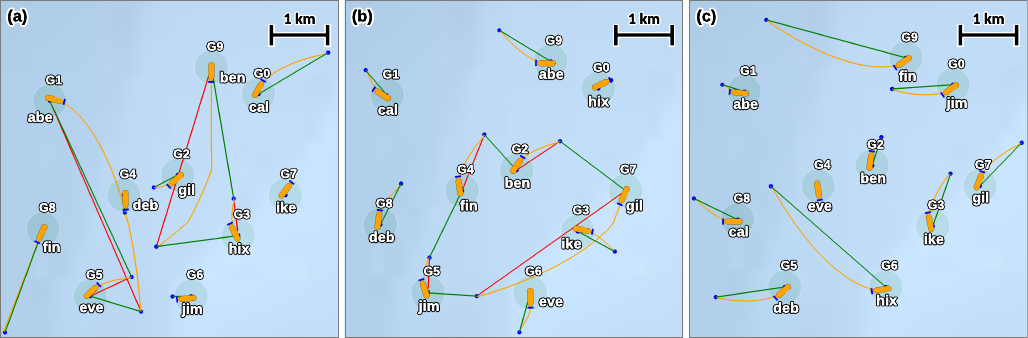}
\caption{Three simulation trails for ten AUVs and ten sample sites using the MOOS-IvP-Multi marine autonomy software.  The maximum communication range $r_0 = 3.5km$ and each vehicle broadcasts a communication of opinion state every $30s$.  The optimality gap was \textbf{a)} 5.0\%, \textbf{b)} 15.7\% \textbf{c)} 0\%. 
Key parameters for all runs: $\gamma = -1.4$, $\kappa = 2$, $\varepsilon= 100$, $\beta_l = \beta_q = \beta_s = 2$, $v_{max}= 1.0$, and all inputs $b_{ij}$ were multiplied by $3e^{-4}$.}
\label{fig:moos_sims}    
\end{figure}

Fig.~\ref{fig:moos_sims} shows three simulation trials with identical sample sites and independently randomized initial AUV locations.  In every trial, all ten AUVs converged to distinct sample sites and the track lines are shown in Fig.~\ref{fig:moos_sims}.  Discrete updates and limited range reduced performance compared to the idealized scenarios considered in earlier sections, with the optimality gap of the emergent allocation from the Hungarian optimum ranging from 4.5\%-16\%, consistent with the other numerical simulation studies, e.g. Fig.~\ref{fig:F1_hist_complete_network_1}.  Overall, these results are encouraging and represent a step toward deployment of the proposed strategy on physical vehicles.

\section{Concluding Remarks}
\label{sec:Concluding Remarks}
We studied a dynamic task allocation model for embodied autonomy that couples agents' movement through space with the dynamics of their opinion variables that reflect preferences towards several tasks. In our simulations, a single social coupling parameter tunes group cohesion, leading to allocation patterns ranging from maximally dispersed task assignments, to intermediate clustered states with several subgroups, to global consensus on a common task. These results show how the same set of local decision rules can support different collective patterns across spatially distributed tasks.

For equal numbers of agents and tasks, we showed through numerical simulations that under strong antagonism the proposed method produces task allocations comparable to those obtained from the Hungarian optimal task assignment algorithm. 
On the other hand, in the cooperative regime, consensus site selection was shown to favor sites that were closest to the group's center of mass. We further illustrated how price capacity feedback can redistribute the macroscopic allocation in the group on the fly, and validated similar behavior under limited communication range and intermittent opinion exchanges in a simulated AUV deployment.
These empirical studies illustrate the promise of opinion-guided autonomy as a flexible, socio-cognitively inspired paradigm for the design of cooperative multi-agent teams, and point to a number of promising directions of future inquiry, including investigating the relationship between spatial geometry, communication structure, and capacity feedback and the feasibility, stability, and optimality of the emergent task allocation dynamics.

\bibliographystyle{splncs04} 
\bibliography{myreferences}

\end{document}